\documentclass[conference,compsoc]{IEEEtran}
\usepackage{amsmath,amssymb}
\usepackage{booktabs,tabularx,array,multirow}
\usepackage{graphicx}
\usepackage{xcolor}
\usepackage{soul}
\soulregister\cite7
\soulregister\ref7
\soulregister\S0
\soulregister\textbf7
\soulregister\emph7
\usepackage{pifont}
\usepackage{algorithm}
\usepackage[noend]{algpseudocode}
\usepackage{cite}
\usepackage[hyphens,spaces]{url}
\usepackage[hidelinks]{hyperref}

\usepackage{tikz}
\usetikzlibrary{arrows.meta}

\usepackage{amsthm}
\theoremstyle{definition}
\newtheorem{definition}{Definition}
\newtheorem{problem}{Problem}
\newtheorem{property}{Property}

\newcommand{\ProjectName}[1]{{\textsc{Precept}}}
\newcommand{\SecProjectName}[1]{{\textsc{Precept}}}
\newcommand{\TabProjectName}[1]{{\textsc{Precept}}}
 
\newif\ifshowresults
\showresultstrue

\newcommand{\E}{\mathbb{E}}
\newcommand{\Ttr}{T_{\mathrm{tr}}}
\newcommand{\Tva}{T_{\mathrm{va}}}
\newcommand{\Tte}{T_{\mathrm{te}}}

\usepackage[normalem]{ulem}   

\newcommand{\best}[1]{\textbf{#1}}
\newcommand{\second}[1]{\underline{#1}}
\newcommand{\hsp}{\begin{tabular}[b]{@{}c@{}}secure-\\pass@1\end{tabular}}

\usepackage{enumitem}

\hypersetup{pdftitle={PRECEPT: Security Self-Improvement of Coding Agents via Evolving Principles}, pdfauthor={Keyan Guo, Rupam Patir, Haipeng Cai, Hongxin Hu}}

\newcommand{\lead}[1]{\par\addvspace{2pt plus 1pt}\noindent\textbf{#1}\enspace\ignorespaces}

\newif\ifcomments
\commentstrue

\newcolumntype{L}{>{\raggedright\arraybackslash}X}
 
\usepackage{tikz}\usetikzlibrary{shapes.geometric}
\definecolor{lgink}{HTML}{1F2937}\definecolor{lgkeep}{HTML}{008300}\definecolor{lgrej}{HTML}{B8BEC8}
\definecolor{lgopen}{HTML}{2A78D6}\definecolor{lgband}{HTML}{DCE8F7}\definecolor{lgref}{HTML}{9CA3AF}\definecolor{lgshade}{HTML}{E3EAF4}\definecolor{lgred}{HTML}{D03B3B}

\begin{document} 

\title{\textsc{Precept}: Security Self-Improvement of Coding Agents via \\Evolving Principles}
 




\author{\IEEEauthorblockN{Keyan Guo}
\IEEEauthorblockA{
University at Buffalo\\
keyanguo@buffalo.edu}
\and
\IEEEauthorblockN{Rupam Patir}
\IEEEauthorblockA{
University at Buffalo\\
rupampat@buffalo.edu}
\and
\IEEEauthorblockN{Haipeng Cai}
\IEEEauthorblockA{
University at Buffalo\\
haipengc@buffalo.edu}
\and
\IEEEauthorblockN{Hongxin Hu}
\IEEEauthorblockA{
University at Buffalo\\
hongxinh@buffalo.edu}
}


%

\maketitle
 
\begin{abstract}

AI coding agents increasingly write and submit code with limited human review, and much of it passes functional tests while remaining insecure. We show that these failures persist. Across 14 task sequences, Codex, Claude Code, and Qwen Code fixed a vulnerability 85.7\% of the time when shown the failing security test, then reintroduced the same vulnerability class in a new session 26.7\% of the time, about as often as agents that had never seen it (27.5\%).
The capability is present; the retention is not. Neither recording the repair in the agent’s persistent context nor supplying published secure coding principles closed the gap: the recorded fix stayed tied to the family of the task it came from, and the principles, written for human developers, remained too general to apply to the code at hand. The knowledge exists; it is not activated. We call the missing ability \emph{security self-improvement}: an agent learns from its own security failures so that each lesson persists across sessions, applies to new code, and does not break what works.

We present \ProjectName{}, which equips coding agents with this ability through an evolving library that starts from authoritative secure coding principles and revises them against the agent's own failures, so that each principle comes to fire on the code where that agent actually goes wrong. A revision is kept only if fresh executions show it makes code more secure while keeping functional success within a small tolerance of the agent without the library. Improvement continues across evolution steps rather than saturating after the first. On task families never seen during evolution, \ProjectName{}'s weakest agent reaches 97.1\% secure working code, above the strongest unaided agent at 81.9\% and the strongest baseline at 87.7\%.

\end{abstract}

\IEEEpeerreviewmaketitle

\section{Introduction}
\label{sec:intro}

AI coding agents are changing how software is developed. Given a feature request, an agent can explore a repository, modify its files, run the tests, and open a pull request on its own~\cite{yang2024sweagent}. On GitHub, the share of pull requests that began with an agent's first draft rose from 17\% to more than 40\% between March and September 2026~\cite{githubOctoverse2026}. The code these agents write, however, often works but is not secure. On real-world feature requests, SWE-agent with Claude~4 Sonnet produced functionally correct code for 57\% of the tasks but correct and secure code for only 11.8\%~\cite{susvibes2026}, and on C/C++ tasks, the best agent evaluated did so for only 23.8\%~\cite{securevibebench2026}. These vulnerabilities arise without any attacker, and 58.2\% of agent-authored pull requests in popular GitHub repositories received no explicit review~\cite{li2025aiteammates}. As developers write and closely review less of the code, its security increasingly depends on the agent that writes it.


What makes this more than a per-task quality problem is that the failures \emph{persist}. We construct 14 task sequences in which an agent writes code that works but fails its security tests, is shown the failing tests and fixes the code, and later meets, in a new session, another task with the same type of vulnerability, as classified by the Common Weakness Enumeration (CWE)~\cite{mitre_cwe} (\S\ref{sec:motivation}). Codex, Claude Code, and Qwen Code fixed the code in 85.7\% of cases, yet on the later task they reintroduced the same CWE in 26.7\% of executions, compared with 27.5\% for a baseline of agents that had never been shown the earlier failure. We call this \emph{vulnerability recurrence}. 
It is a failure of retention rather than of capability, because when the fix stayed in context, recurrence on a variant of the fixed task fell from 36.7\% to 13.9\%.
Left to themselves, these agents did not learn from their security failures. A developer whose vulnerability is pointed out in review is expected to avoid it in later work, but an autonomous workflow may have no such developer, so the agent itself must keep the lesson.

Two remedies are available to an agent, and neither prevents recurrence on its own. The first is the agent's persistent context, the instruction files and memories that it reads in every session. In our study (\S\ref{sec:motivation}), no agent recorded a fix there on its own, and the notes the agents wrote when asked often described the specific task they had fixed rather than a general lesson, so they helped on variants of that task but much less on other tasks with the same CWE.
The second is published secure coding principles, which codify the knowledge needed to avoid these vulnerabilities~\cite{owasp_scp,owasp_asvs,owasp_cheatsheets,openssf_python,cert_coding}. Written for human developers, they state what should hold in general and leave it to the reader to decide where it applies. Given to the agents, neither the complete set nor only the relevant principles measurably lowered recurrence, whereas the relevant principle restated for the class of code at hand did, suggesting that the knowledge was available but not applied where the agents failed.

Both findings point to the same need. \emph{A coding agent must learn from its own security failures}, which reveal where its knowledge is not retained or not applied. Prior work makes AI-generated code more secure by training or steering the model, constraining its decoding, or retrieving security knowledge for each task~\cite{sven2023,safecoder2024,fu2024constrained,lin2025codeguarder,mamun2026secgent,patir2025grasp}, but the security knowledge these methods rely on does not change with what the agent gets wrong. AI agents have been shown to learn from their own experience, revising their memories, skills, workflows, contexts, or rules~\cite{shinn2023reflexion,yang2026skillopt,wang2025awm,zhang2026agentic,zhao2024expel,jiang2026ruleevolve}, and work on recursive self-improvement (RSI) has agents repeatedly improve their own code~\cite{zelikman2024stop,robeyns2025sica,zhang2026dgm} or specifications~\cite{cheng2026specforge}. These methods learn toward task success or efficiency, and under that objective a vulnerability rarely counts as a failure, because vulnerable code passes its functional tests. Coding agents need an analogous ability for the security of the code they write. We call it \textbf{security self-improvement}, in which a coding agent learns from its own security failures so that each lesson persists across sessions, applies to new code, and does not break what works.

Security self-improvement can take different forms, and we pursue it through secure coding principles~\cite{owasp_scp}. Our study found that a published principle became measurably helpful once restated for a class of code, so learning can start from the published principles and evolve them from the agent's security failures. Principles also suit the requirements of security self-improvement. Kept in project instructions, principles remain available across sessions without changing model weights. Because they describe a class of code, their lessons can apply to new code beyond an individual fix. As separate units of text, principles can be revised and tested to improve security without breaking what works, and they remain readable for developers.

Evolving principles from an agent’s own failures raises three challenges. The first is \emph{attribution}. A failure shows that the code is insecure or broken, not which part of the library failed: a principle may be missing, unread, misapplied, or applied so strictly that valid input is rejected, and each calls for a different change. The second is \emph{generalization}. A lesson drawn from one failure tends to fit only its task, as the agents’ own notes did (\S\ref{sec:motivation}), while a lesson stated too generally goes unapplied, as the published principles did. The third is \emph{estimation}. A revision’s effect can be smaller than the agent’s run-to-run variation, and security tests alone cannot tell secure code from code that no longer works~\cite{cweval}, so a revision can appear to help by chance or by making code refuse its input.

We present \ProjectName{} (\textbf{P}reventing \textbf{R}ecurring \textbf{E}rrors in \textbf{C}oding by \textbf{E}volving \textbf{P}rinciples across \textbf{T}asks), which equips coding agents with security self-improvement. For each agent, \ProjectName{} evolves a library of principles that the agent reads on every task, starting from the published principles and changing them only through a closed set of mutation primitives, so each principle can be traced to its source and each change to the step that made it. Each evolution step executes the agent on the training tasks (\textsc{Test}) and then takes the three challenges in turn. \textsc{Analyze} attributes each failure, from the agent’s executions, to the part of the library that failed, and that determines which primitives may fix it. \textsc{Draft} writes each revised principle for a class of code, stating where it applies, what must hold there, and what must remain allowed, without the names or details of any training task. \textsc{Decide} keeps a revision only when fresh, paired executions, pooled across steps, show that it helps while functional code stays above a fixed floor. Acceptance rests only on executions and their statistics, not on a model's judgment.

We apply \ProjectName{} to three coding agents, Codex, Claude Code, and Qwen Code, and evaluate it on two secure code generation benchmarks, SeCodePLT~\cite{nie2026secodeplt} and CWEval~\cite{cweval}, with test tasks held out from training. 
With \ProjectName{}, 97.1--98.5\% of the agents' functional code on SeCodePLT is secure, compared with 81.9\% for the strongest agent on its own and 87.7\% for the strongest baseline. The evolved libraries raise the rate of code that is both functional and secure by 8.9--11.6 percentage points over the agents on their own, 7.0--8.1 points over the published principles, and 6.9--9.7 points over each agent's best learned baseline~\cite{zhang2026agentic,yang2026skillopt,jiang2026ruleevolve}, and all three agents produce functional code at least as often as on their own. In new sessions on the chains of our motivating study, recurrence falls from 26.7\% after the fix alone to 11.5\% with \ProjectName{}, compared with 15.0\% when the agents carry forward their own notes. 
Improvement continues as evolution proceeds, and ablations show the role of each component. The evolved libraries also transfer across agents and models, where they still significantly raise the share of functional code that is secure.

This paper makes the following contributions:
\begin{itemize} 
\item We identify vulnerability recurrence, in which coding agents reintroduce a CWE they have previously introduced and fixed, and show that it is a failure of retention that neither persistent context nor published secure coding principles reliably prevents (\S\ref{sec:motivation}).
\item We define security self-improvement as a property of a learning procedure, in which an agent learns from its own security failures so that each lesson persists across sessions, applies to new code, and does not break what works (\S\ref{sec:problem-formulation}).
\item We present \ProjectName{}, which evolves a library of principles from an agent's own failures through traceable mutation primitives, attributes each failure to the part of a principle that failed, and keeps a revision only when fresh, paired executions show that it helps while functional code stays above a floor (\S\ref{sec:method}).
\item We evaluate \ProjectName{} with three coding agents and show that it lowers recurrence across sessions from 26.7\% to 11.5\%, raises the share of functional code that is secure on unseen task families from 79.7--81.9\% to 97.1--98.5\%, keeps improving security as evolution proceeds, and evolves libraries that transfer across agents and models (\S\ref{sec:evaluation}).

\end{itemize}


\section{Motivating Study}
\label{sec:motivation}

This study examines whether coding agents learn from their security failures and, if they do not, whether the remedies available to them close the gap. We follow each failure forward in time. After an agent fixes a vulnerability, we observe whether it avoids vulnerabilities of the same CWE on later tasks, with and without these remedies.

\subsection{Research Questions}

\begin{itemize}[leftmargin=1em, itemsep=1pt]
    \item \textbf{RQ1~(Recurrence):} After fixing a vulnerability, do coding agents introduce the same type of vulnerability again on later tasks?
    \item \textbf{RQ2~(Capability vs.\ retention):} Is recurrence a failure of security capability or a failure of retention?
    \item \textbf{RQ3~(Sufficiency of persistent context):} Does carrying what an agent learned from an earlier failure into later sessions prevent recurrence?
    \item \textbf{RQ4~(Sufficiency of published principles):} Does supplying published secure coding principles, either a complete set or only those relevant, prevent recurrence?
\end{itemize}

\subsection{Study Design}
\label{sec:motivation-setup}

\lead{Sessions and persistent context.} A \emph{session} is one conversation with a coding agent and may span several tasks. An \emph{execution} is the agent's work on one task, a sequence of \emph{turns}, each one a model call and the tool calls it makes. What carries over to the next session is the agent's \emph{persistent context}, its instruction files, such as \texttt{AGENTS.md}~\cite{agentsmd} or \texttt{CLAUDE.md}~\cite{claudemd}, and its memory.\label{session_pc}

\lead{Agents and tasks.} We study Codex, Claude Code, and Qwen Code, all running on the same underlying model, Qwen3.6 (\S\ref{sec:setup}), on Python tasks from SeCodePLT~\cite{nie2026secodeplt}. Each task targets one CWE and comes with functional and security tests, and we count a vulnerability only when the security tests detect it. SeCodePLT writes each coding problem in several variants, and we call the variants of one problem a \emph{task family}.

\lead{Chains.} Each task sequence, or \emph{chain}, starts from a task $T_i$ on which all three agents wrote functional but insecure code. Each agent is shown its failing security tests, with no vulnerability named, and asked to fix its code, and the chain continues only if the fix passes both kinds of tests. The agent then works on $m \in \{0, 3, 6\}$ tasks with other CWEs and then meets two later tasks $T_j$ with the same CWE as $T_i$, one from $T_i$'s task family and one from a different family. We take the chains from task families held out from training, choosing as $T_i$ the tasks that all three agents failed most often, at most two per CWE. This gives 14 chains over 12 of SeCodePLT's 17 CWEs. A chain needs a starting task on which every agent introduces the vulnerability, so that each has something to fix. For the other five CWEs, no held-out task qualified: on each, at least one agent wrote secure code or code that did not work.

\lead{Conditions.} The agent meets $T_j$ (1) with no earlier failure and fix (\emph{baseline}); (2) in a new session, which carries over only its persistent context, so it recalls the fix only if it saved it (\emph{new session}); (3) in the session that continues the conversation of the fix (\emph{same session}); or (4) in a new session after writing what it learned into its instruction file when asked (\emph{own note}). A preliminary study of 135 fixes with $m = 0$ adds a \emph{verbatim record}, the failed code and its fix pasted into the next session's task prompt (Appendix~\ref{app:motivation}).

\lead{Published principles.} We also execute each $T_j$ in a new session without an earlier fix, with \emph{guidance}, text given to the agent to shape the code it writes, added to its prompt at six levels of specificity (Table~\ref{tab:ladder} in Appendix~\ref{app:motivation}): (a) a one-line security reminder~\cite{susvibes2026}; (b) all 218 principles of the OWASP Secure Coding Practices~\cite{owasp_scp}; (c) the principles relevant to $T_j$'s CWE, which we select by hand as an oracle retrieval; (d) the relevant principle restated for the class of code, naming where it applies and what must hold without naming the task; (e) guidance written for $T_j$ itself, which says what its code must check without describing the reference fix; (f) a description of what the benchmark's reference fix does.

\lead{Measurement.} Every later task is executed five times per condition and is compared with the baseline, paired by task, with 95\% bootstrap intervals over chains. 


\subsection{Metrics}
\label{sec:motivation-metrics}

Let $V(T) = 1$ when an execution of task $T$ introduces the vulnerability, that is, when its code passes the functional tests but fails the security tests.  

\begin{itemize}[leftmargin=1em, itemsep=1pt]
    \item \textbf{(BIR) Baseline incidence rate.} $\mathrm{BIR} = \Pr[V(T_j) = 1]$ in the baseline, how often the agent introduces the vulnerability without an earlier fix.
    \item \textbf{(RR) Recurrence rate.} $\mathrm{RR} = \Pr[V(T_j) = 1 \mid \text{fix at } T_i]$, how often the agent introduces the vulnerability again after fixing it.
    \item \textbf{(RG) Retention gain.} $\mathrm{RG} = \mathrm{BIR} - \mathrm{RR}$, how much less often the vulnerability appears because of the earlier fix. It is 0 when the fix leaves no trace and equals BIR when the vulnerability never returns.
    \item \textbf{(FR) Fix rate.} The share of executions at $T_i$ that fix the vulnerability without losing functional correctness. It shows whether the agent can write the secure version.
    \item \textbf{{(CC) Catalog coverage.}} CC is the share of later tasks whose vulnerability a principle addresses directly.
    
\end{itemize}

\begin{table}[t]
\centering
\caption{Recurrence in a new session (\%).}
\label{tab:recurrence}
\begin{tabular}{lcccc}
\toprule
Agent & FR$\uparrow$ & BIR$\downarrow$ & RR$\downarrow$ & RG$\uparrow$ \\
\midrule
Codex & 78.6 & 30.0 & 24.8 & $+5.2$ \\
Claude Code & 85.7 & 27.5 & 25.6 & $+1.9$ \\
Qwen Code & 92.9 & 25.4 & 29.2 & $-3.8$ \\
All & 85.7 & 27.5 & 26.7 & $+0.8$ \\
\bottomrule
\end{tabular}
\end{table}

\subsection{Results}
\label{sec:motivation-results}

\lead{RQ1: Yes, agents reintroduce the same CWE after fixing it.} In a new session, the agents introduced the same CWE at the later tasks in 26.7\% of executions, averaged over both later tasks and all three distances from the fix, against 27.5\% with no earlier fix (Table~\ref{tab:recurrence}), a retention gain whose interval includes zero. The gain stays small for every agent, distance, and family (Table~\ref{tab:chains}). A fix thus leaves almost no trace on later work, and an agent that has fixed a vulnerability is about as likely to write it again as one that never met it.

\begin{table}[t]
\centering
\caption{Recurrence at the later task after a fix (\%).}
\label{tab:chains}
\resizebox{\columnwidth}{!}{%
\begin{tabular}{llcccc}
\toprule
& & \multicolumn{2}{c}{Same family (BIR 36.7)} & \multicolumn{2}{c}{Other family (BIR 18.3)} \\
\cmidrule(lr){3-4}\cmidrule(l){5-6}
Condition & $m$ & RR$\downarrow$ & RG$\uparrow$ & RR$\downarrow$ & RG$\uparrow$ \\
\midrule
New session & 0 & 35.6 & $+1.1$ & 21.7 & $-3.3$ \\
 & 3 & 35.8 & $+0.6$ & 18.9 & $-0.6$ \\
 & 6 & 30.6 & $+6.1^{*}$ & 17.8 & $+0.6$ \\
\midrule
Same session & 0 & 13.9 & $+22.8^{*}$ & 20.0 & $-1.7$ \\
 & 3 & 17.1 & $+19.4^{*}$ & 24.0 & $-5.1$ \\
 & 6 & 15.4 & $+21.1^{*}$ & 25.1 & $-6.3$ \\
\midrule
Own note & 0 & 13.9 & $+22.8^{*}$ & 13.9 & $+4.4$ \\
 & 3 & 14.4 & $+22.2^{*}$ & 15.6 & $+2.8$ \\
 & 6 & 15.6 & $+21.1^{*}$ & 16.7 & $+1.7$ \\
\bottomrule
\end{tabular}%
}
\par\smallskip\raggedright\footnotesize
$^*$: retention gain whose 95\% bootstrap interval over chains lies above 0.
\end{table}

\begin{table}[t]
\centering
\caption{Published principles at the later task (\%).}
\label{tab:ladder_result}
\begin{tabular}{lcc}
\toprule
Guidance & RR$\downarrow$ & RG$\uparrow$ \\
\midrule
None (BIR) & 27.5 & -- \\
(a) Security reminder & 25.3 & $+2.2$ \\
(b) All 218 OWASP principles & 24.2 & $+3.3$ \\
(c) Relevant OWASP principles & 23.3 & $+4.2$ \\
(d) Class-specific principle & 16.7 & $+10.8^{*}$ \\
(e) Guidance written for the task & 6.9 & $+20.6^{*}$ \\
(f) Reference-fix description & 8.1 & $+19.4^{*}$ \\
\bottomrule
\end{tabular}
\par\smallskip\raggedright\footnotesize
$^{*}$: positive gain with 95\% confidence.
\end{table}

\lead{RQ2: Recurrence is a failure of retention, not of capability.} The agents fixed 85.7\% of the vulnerabilities (Table~\ref{tab:recurrence}), and while the fix stayed in context (same session), recurrence on a later task of the same family fell by about 20 points and stayed lower six tasks later (Table~\ref{tab:chains}). The agents can therefore write the secure version and apply it to a later variant of the task when the fix remains in the conversation. However, the fix did not lower recurrence on another family of the same CWE, which suggests that what the agents keep is tied to the fixed task rather than to the CWE.

\lead{RQ3: Persistent context helps only when the agent is asked, and only on the same family.} With memory features enabled, no agent recorded the fix, and none of the 36 later sessions held a trace of it. When asked, the agents wrote a note in 28 of 36 cases, and the note lowered recurrence on the same family as much as keeping the fix in context did, but not on another family (Table~\ref{tab:chains}). Nearly half of the notes (13 of 28) tie the lesson to the fixed task's own functions or fields, and a verbatim record of the failed code and its fix was likewise tied to the fixed task's family in the preliminary study (Appendix~\ref{app:motivation}). Persistent context can carry a fix forward, but agents do not use it for that on their own, and what they write describes the task they fixed rather than a lesson for the CWE.

\label{sec:c1}
\lead{RQ4: No. Published principles cover most of these vulnerabilities but do not prevent them.} An OWASP principle directly addresses the vulnerability in 19 of the 27 later tasks (CC = 70\%; a second rater agrees on 25 of the 27, Cohen's $\kappa = 0.83$, Appendix~\ref{app:motivation}), yet neither all 218 principles nor only the relevant ones measurably lower recurrence (levels (b) and (c) in Table~\ref{tab:ladder_result}). The knowledge these tasks need is largely published, but in its general form it does not produce the checks that the code requires. 
Recurrence falls as the guidance becomes more specific. The same knowledge restated for the class of code (level (d)) lowers it by 10.8 points with little observed loss of working code, and guidance that states what the task's code must check (level (e)) is about as effective as describing the reference fix (level (f)). Levels (e) and (f) serve only as reference points, however. Each must be written for a single task by someone who already knows its vulnerability, and it does not carry over to other tasks, so supplying such guidance to an agent across many tasks is impractical.

\section{Threat Model and Problem Statement}
\label{sec:problem}

\subsection{Threat Model}
\label{sec:threat-model}

\lead{The agent, not an attacker, introduces the vulnerability.} The agent works on ordinary coding tasks that no attacker controls and means to meet their requirements, so every vulnerability is its \emph{own error} rather than one planted to provoke it. It may lack the security knowledge it needs, overlook it, misapply it, or apply it so strictly that valid input is rejected.

\lead{The attacker exploits deployed code.} Once deployed, a vulnerability that the agent wrote can be exploited through the software's inputs, letting an attacker, for example, exhaust resources (CWE-770), alter the commands the code issues (CWE-77), or expose sensitive data (CWE-200). Each task's security tests check whether its implementation withstands the attacks that its CWE describes, and we call code ``secure'' when it passes them, so our claims about security are limited to those attacks.


\lead{Tasks, tests, and the learning process are trusted.} Our focus is how an agent learns from its own security failures. We therefore assume that no adversary controls the task requirements, the tests, the recorded execution feedback, or the learning infrastructure, and changing the tests is never an admissible way to improve a score. Section~\ref{sec:disc-robust} discusses adversaries who target this learning or the text the agent reads at deployment.

\subsection{Problem Formulation}
\label{sec:problem-formulation}

A coding agent receives a development task $x$, made of functional requirements and a software environment. We split the agent into a fixed part $\mathcal{A}$ and a part $B$ that learning may change, such as its memories, skills, instructions, scaffold code, or model weights. With execution randomness $\xi$, it produces an implementation $c_B$ and a trajectory $\tau_B$. We write $f_x(c), s_x(c) \in \{0, 1\}$ for whether $c$ passes the functional and security tests, and $j_x(c) = f_x(c)\,s_x(c)$ for success. An execution without a testable implementation counts as neither functional nor successful.

\lead{The agent that improves.} A learning procedure $L$ updates $B$ from the agent's own executions, $B_{t+1} = L(B_t, E_t)$, where $E_t$ holds the trajectories and test outcomes of the agent with $B_t$ on the training tasks, where a test outcome names each failing test case and, where the benchmark gives them, its inputs, the expected behavior, and what the code did, but not the test code. Where the benchmark has them, $L$ may also read the training tasks' reference fixes. Starting from the agent's initial $B_0$, it produces a sequence $B_0, B_1, \dots$, and security self-improvement asks that this sequence make the agent more secure on tasks it has not yet seen without breaking what works.

\lead{What learning requires.} Learning draws on the agent's executions on past tasks and serves its future tasks. It assumes that the tasks it learns from come with functional and security tests. Such tests accumulate in practice as vulnerabilities are found and fixed, since a fix often comes with a test that reproduces the attack. We consider learning that updates $B$ from a batch of tasks at once rather than after each failure, because the effect of a change on a single task cannot be told apart from the agent's run-to-run variation~\cite{miller2024errorbars}. $B$ is learned for each agent and model. Tasks are drawn from a distribution $\mathcal{X}$ over task families, and the training tasks $\Ttr$, which stand for the past, the validation tasks $\Tva$, and the test tasks $\Tte$, which stand for the future, come from disjoint sets of families. Choices made during learning, such as which $B_t$ to return, use $\Tva$ alone.
\lead{Rates, not single outcomes.} Because the agent is stochastic, $B$ changes the probability of each outcome. Over tasks $x \sim \mathcal{X}$ and randomness $\xi$, three rates describe $B$,
\begin{equation}
\begin{aligned}
J(B) &= \E_{x}\,\E_{\xi}\big[j_x(c_B)\big],\\
P(B) &= \E_{x}\,\E_{\xi}\big[f_x(c_B)\big],\\
K(B) &= \E_{x}\,\E_{\xi}\big[\mathrm{tok}(\tau_B)\big],
\end{aligned}
\label{eq:jpk}
\end{equation}
where $\mathrm{tok}(\tau_B)$ is the number of input tokens consumed. $J$ is secure-pass@1~\cite{fu2024constrained}, $P$ is pass@1, and $S_{|P} = J/P$ is the share of functional implementations that are also secure, pooled over all of them. We optimize $J$ rather than the rate of code that passes the security tests, because code that rejects every input passes them, while $J$ counts security only for code that works~\cite{cweval,dai2026rethinking}. We report $S_{|P}$ together with $P$, since $S_{|P}$ alone rises when insecure code stops working.

\begin{definition}[Security self-improvement]
\label{def:ssi}
A learning procedure $L$ gives an agent security self-improvement if $L$ and its sequence $B_0, B_1, \dots$ meet four conditions.
\begin{itemize}[leftmargin=1em, itemsep=1pt]
\item \emph{Own experience.} $L$ revises $B$ only with what the training tasks provide, the agent's own executions, their test outcomes, and the tasks' reference fixes where they exist. It receives no guidance written for particular tasks and nothing from the test tasks.
\item \emph{Persistence.} What the agent learns from a failure continues to lower how often it introduces the same CWE in later sessions ($\mathrm{RG} > 0$, \S\ref{sec:motivation-metrics}).
\item \emph{New code.} The $B_t$ that learning returns raises $J$ above $J(B_0)$ on task families never seen in learning.
\item \emph{No regression.} $P(B_t) \ge P(B_\varnothing) - \epsilon_F$ for every $t$, where $B_\varnothing$ is the unmodified agent and $\epsilon_F$ is a small tolerance.
\end{itemize}
\end{definition}

\lead{Objective.} Learning seeks the $B$ that maximizes $J$ subject to the no-regression floor of Definition~\ref{def:ssi},
\begin{equation}
B^\star \in \arg\max_{B\in\mathcal F} J(B), \quad \mathcal F = \{B : P(B) \ge P(B_\varnothing) - \epsilon_F\}.
\label{eq:objective}
\end{equation}
Eq.~\ref{eq:objective} encodes two conditions of Definition~\ref{def:ssi}, new code through $J$ on unseen task families and no regression through $\mathcal{F}$. The other two lie outside the objective. Own experience limits what learning may use, and persistence is measured with the chains of the motivating study (\S\ref{sec:persist}). The floor is the functional level of the unmodified agent rather than that of the current $B$, so that small losses cannot accumulate, and we set $\epsilon_F = 3$ points, above what the one-line security reminder costs: it changes pass@1 by $-1.8$ to $+1.8$ points on SeCodePLT's test tasks (Table~\ref{tab:secodeplt}). Among candidates with equal $J$, the one with the lower $K$ is preferred.

\section{\ProjectName{} Design}
\label{sec:method}

\begin{figure}[t]
\centering
\begin{tikzpicture}[font=\footnotesize, >={Stealth[length=1.5mm]}]
\tikzset{
  stage/.style={draw, thick, rounded corners=2pt, fill=gray!10, minimum width=13.5mm, minimum height=10mm, align=center, inner sep=1pt},
  lab/.style={font=\scriptsize, align=center, inner sep=1pt},
  num/.style={circle, fill=black, text=white, font=\tiny\bfseries, inner sep=0pt, minimum size=3.2mm}
}
\foreach \d in {3pt,1.5pt}
  \draw[thick, fill=white, rounded corners=1.5pt] ([shift={(-\d,\d)}]-0.6,-0.5) rectangle ([shift={(-\d,\d)}]0.6,0.5);
\node[draw, thick, fill=white, rounded corners=1.5pt, minimum width=12mm, minimum height=10mm, align=center, inner sep=1pt] (lib) at (0,0) {Library $B_t$\\[-1pt]{\scriptsize $(I, \mathcal{D})$}};
\node[stage, fill=red!10] (test)    at (1.8,0) {\textsc{Test}\\[-1pt]{\scriptsize records $E$}};
\node[stage, fill=orange!18] (analyze) at (3.4,0) {\textsc{Analyze}\\[-1pt]{\scriptsize causes $\gamma$}};
\node[stage, fill=blue!10] (draft)   at (5.0,0) {\textsc{Draft}\\[-1pt]{\scriptsize primitives $\delta$}};
\node[stage, fill=green!14] (decide)  at (6.6,0) {\textsc{Decide}\\[-1pt]{\scriptsize fresh, paired}};
\foreach \n/\s in {1/test,2/analyze,3/draft,4/decide}
  \node[num] at (\s.north west) {\n};
\node[draw, thick, cylinder, shape border rotate=90, aspect=0.22, fill=gray!12, minimum width=12.5mm, minimum height=7mm, inner sep=1.5pt] (ledger) at (5.8,-1.4) {ledger $\mathcal{H}$};
\node[num] (n5) at (0.45,-1.4) {5};
\node[lab, anchor=west] (sel) at ([xshift=1pt]n5.east) {$B^\star$ selected on $\Tva$};
\draw[->, thick] (lib.east) -- node[lab, above] {$\Ttr$} (test.west);
\draw[->, thick] (test) -- (analyze);
\draw[->, thick] (analyze) -- (draft);
\draw[->, thick] (draft) -- (decide);
\draw[->, thick] (decide.north) -- ++(0,0.5) -| node[lab, pos=0.25, above] {accept: $B_{t+1} = B_t \oplus \delta$} ([yshift=3pt]lib.north);
\draw[->, thick] (decide.south) |- node[lab, pos=0.3, right] {reject} (ledger.east);
\draw[->, thick] (ledger.west) -| node[lab, pos=0.75, left] {retry} (draft.south);
\draw[->, thick] (lib.south) |- (n5.west);
\end{tikzpicture}
\caption{The evolution loop of \ProjectName{}.}
\label{fig:overview}
\end{figure}
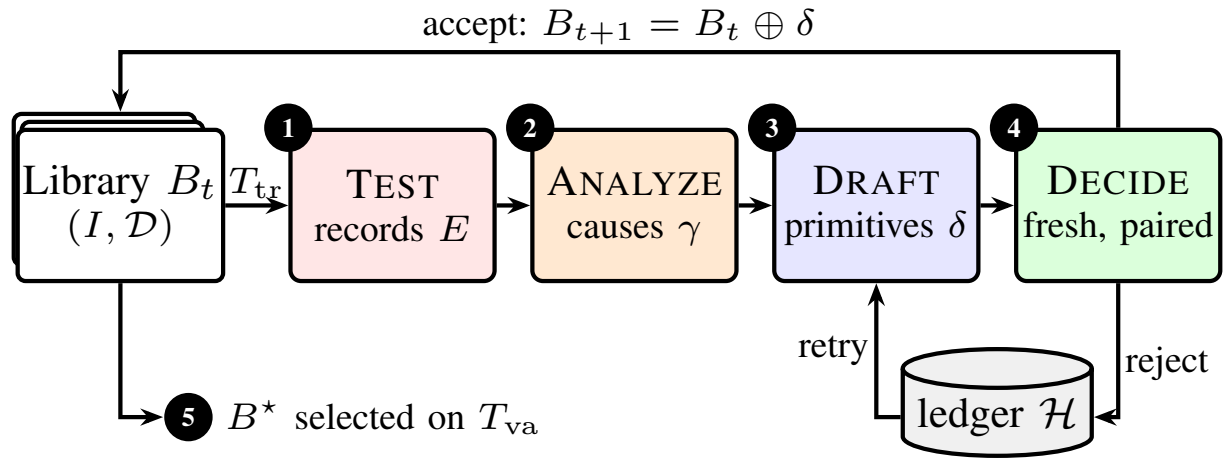

\subsection{Overview}
\label{sec:design}
\label{sec:rationale}
\label{sec:requirements}


\begin{figure*}[t]
\centering
\includegraphics[width=0.95\textwidth]{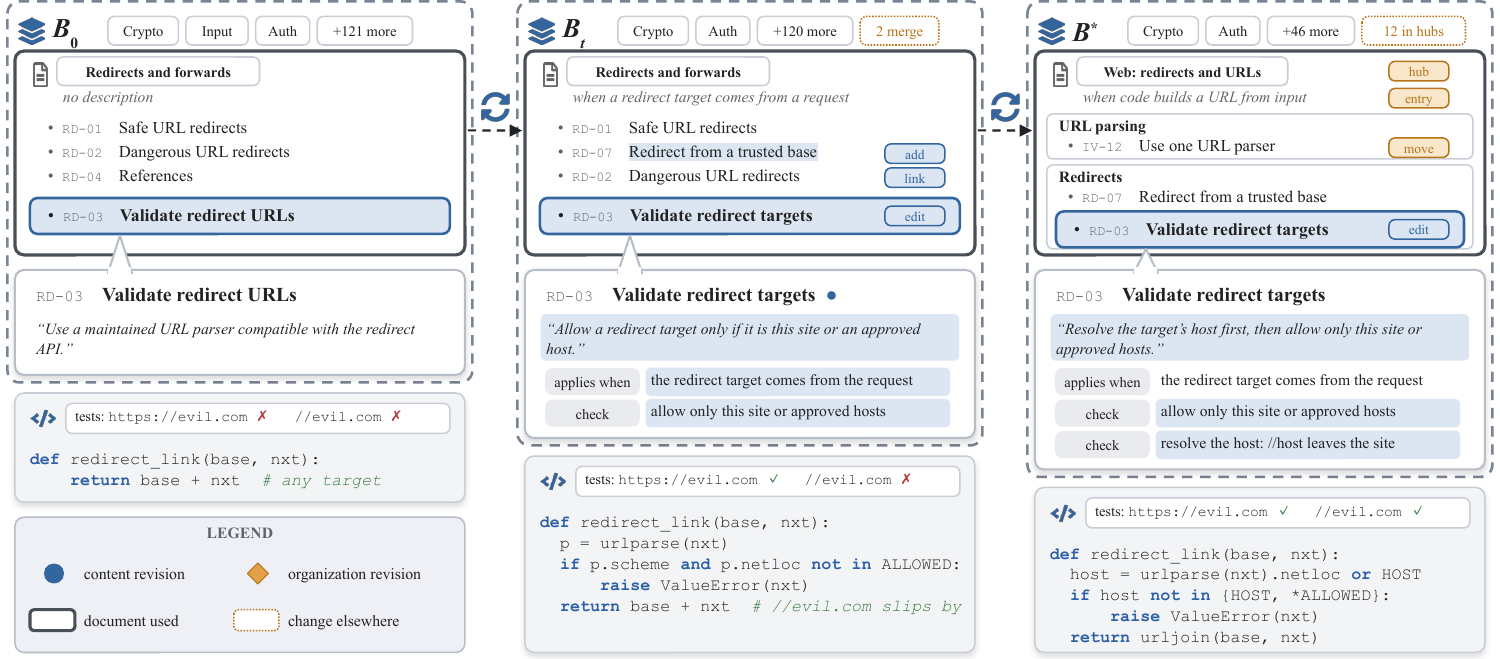}
\caption{Evolution of a redirect-validation principle (dashed boxes) and the agent's code with each library on one held-out task (grey boxes).}
\label{fig:principle-evolution}
\end{figure*}


Definition~\ref{def:ssi} applies to any part $B$ of the agent that learning changes, such as the model's weights or the context, skills, or rules the agent reads~\cite{zhang2026agentic,yang2026skillopt,jiang2026ruleevolve}. We take $B$ to be a library of secure coding principles, with the model and scaffold fixed. The agent reads the library in every session, so lessons persist without changing the model. A principle covers a class of code, and a principle restated for a class of code lowered recurrence where the published principles did not (\S\ref{sec:motivation-results}). Published principles already cover most of these vulnerabilities, so learning can start from them, and every change remains an auditable text. The rest of this section develops how the library is revised.

Given a coding agent, the published principles, and training and validation tasks, \ProjectName{} searches for the library $B^\star$ of Eq.~\ref{eq:objective}, changing the published principles only through mutation primitives (\S\ref{sec:guidebook}, \S\ref{sec:ops}). The objective has no gradient, so each candidate is judged by the agent's executions (\S\ref{sec:optimization}), which raises three problems. An execution shows that code failed but not which part of the library to change (\emph{attribution}, \S\ref{sec:diagnose}). A revision drafted from training failures must also help on unseen task families (\emph{generalization}, \S\ref{sec:draft}). Whether it helps must be decided from a few executions that vary from run to run, without letting functional code fall (\emph{estimation}, \S\ref{sec:decide}). Each step takes these problems in turn (Figure~\ref{fig:overview}). \textsc{Test} executes the agent with the current library on the training tasks, \textsc{Analyze} attributes its failures, \textsc{Draft} writes a revision for a class of code, and \textsc{Decide} accepts it or keeps it in the ledger for a later retry. The final library is chosen on the validation tasks.

\subsection{Principles and the Library}
\label{sec:guidebook}
\label{sec:example}

As a running example, Figure~\ref{fig:principle-evolution} shows one principle for validating redirect targets in the seed, midway through a run, and in the final library, together with the code the agent writes with each library on the same held-out task.

\begin{definition}[Principle]
\label{def:principle}
A principle is a tuple $\pi = (\mathit{id}, s, a, k, v, x, \rho)$ of an identifier $\mathit{id}$, a statement $s$, the situation $a$ in code to which it applies, the check $k$ that must hold there, the behavior $v$ that must remain allowed, supporting fields $x$ such as details, examples, and links to related CWEs and principles, and a provenance $\rho$ that records the published source the principle instantiates and the step that last changed it. Only $s$ is required, and any other field may be empty ($\bot$).
\end{definition}

Let $\Sigma(c)$ be the situations in an implementation $c$, such as each point where caller-supplied data is stored. A principle $\pi$ holds for $c$ when, in every situation its $a$ describes, the check $k$ holds and the behavior $v$ remains allowed,
\begin{equation}
c \models \pi \iff \forall \sigma \in \Sigma(c):\ a(\sigma) \Rightarrow \big(k(\sigma) \wedge v(\sigma)\big).
\label{eq:principle}
\end{equation}
We call the situations where $a$ matches the principle's \emph{scope}, which is all code when $a = \bot$. Because the fields are natural language, no automated tool such as a static analyzer evaluates Eq.~\ref{eq:principle}. The agent judges whether a situation matches and the check holds, so the equation states what a principle asks of the code, not how it is enforced.


A published principle typically gives only its statement, so its scope is all code, and $a$, $k$, and $v$ are left to the reader, the gap our motivating study found (\S\ref{sec:c1}). In Figure~\ref{fig:principle-evolution}, principle RD-03 in the seed $B_0$ has only its statement, and the agent's code redirects to any target. At the other extreme, a principle written for one task has that task's scope and does not transfer. Evolution aims between the two, at a principle whose scope is a class of code, with $k$ precise enough to act on and $v$ explicit enough that valid input is still accepted.

\begin{definition}[Library]
\label{def:library}
A library $B = (I, \mathcal{D})$ consists of a set of documents $\mathcal{D}$, each holding principles, and an index $I$ whose entries name a document, state what it covers, and say when to open it, preceded by a short reading procedure. Every library in a run satisfies two invariants. Every principle is reachable from $I$, and $B$ changes only through the primitives of \S\ref{sec:ops}.
\end{definition}

The agent receives only $I$ and opens a document when it judges that the document applies (\S\ref{sec:setup}), so a principle can fail through what it says or through not being found.


\lead{Initialization.} Evolution starts from a seed library $B_0$ built from OWASP's verification standard, Secure Coding Practices, and cheat sheets~\cite{owasp_asvs,owasp_scp,owasp_cheatsheets} and from language-specific guides and static analyzers (Appendix~\ref{app:impl}). Each published principle enters the seed verbatim in the form of Definition~\ref{def:principle}, with its source as provenance and $a$, $k$, and $v$ empty unless the source states them, and the principles of each source chapter form one indexed document. Apart from the index's reading procedure, the seed contains no text of ours, so everything an evolved library holds beyond the seed was added by evolution. In a preliminary run, a model-written seed scored below no guidance by teaching the agent to reject valid input.
The floor of Definition~\ref{def:ssi} is the agent without a library, $B_\varnothing$, not $B_0$.

\subsection{Mutation Primitives}
\label{sec:ops}

A revision should change only what the evidence calls for. A language model asked to rewrite a document can drop or reword text it was not asked to change, and over many steps such drift would erode the published principles unnoticed. \ProjectName{} therefore changes the library only by composing a closed set of mutation primitives $\mathcal{O}$ (Table~\ref{tab:ops}).

\begin{definition}[Revision]
\label{def:revision}
A mutation primitive $o \in \mathcal{O}$ changes one named target, a principle, a field, a document, or an index entry, and applies only if the target exists and, for an edit, still holds the text being replaced. A revision $\delta = (o_1, \dots, o_\ell)$ applies its primitives in order, $B^{(i)} = o_i(B^{(i-1)})$ from $B^{(0)} = B$, and $B \oplus \delta = B^{(\ell)}$ is defined only if each one applies.
\end{definition}

\begin{table}[t]
\centering
\caption{Mutation primitives. A field $f$ is one of $s$, $a$, $k$, $v$, $x$.}
\label{tab:ops}
\footnotesize
\setlength{\tabcolsep}{4pt}
\begin{tabular*}{\columnwidth}{@{\extracolsep{\fill}}@{}ll@{}}
\toprule
Primitive & Effect \\
\midrule
\multicolumn{2}{@{}l}{\emph{Content}} \\
\hspace{1em}$\mathrm{add}(d, \pi)$ & add $\pi$ to document $d$ \\
\hspace{1em}$\mathrm{edit}(\pi, f, t)$ & set field $f$ of $\pi$ to text $t$ \\
\hspace{1em}$\mathrm{split}(\pi)$, $\mathrm{join}(\pi)$ & separate or rejoin its details \\
\hspace{1em}$\mathrm{merge}(\pi, \pi')$ & combine two principles into one \\
\hspace{1em}$\mathrm{link}(\pi, \pi')$, $\mathrm{unlink}(\pi, \pi')$ & add or remove a relation \\
\midrule
\multicolumn{2}{@{}l}{\emph{Organization}} \\
\hspace{1em}$\mathrm{create}(d)$, $\mathrm{rename}(d)$, $\mathrm{delete}(d)$ & manage a document \\
\hspace{1em}$\mathrm{move}(\pi, d)$ & move $\pi$ to document $d$ \\
\hspace{1em}$\mathrm{entry}(d, t)$ & edit the index entry of $d$ \\
\hspace{1em}$\mathrm{hub}(d, \mathcal{D}')$ & group documents $\mathcal{D}'$ under $d$ \\
\hspace{1em}$\mathrm{procedure}(t)$ & set the reading procedure \\
\bottomrule
\end{tabular*}
\end{table}

\emph{Content} primitives change the fields of principles, and \emph{organization} primitives change the index and the grouping of principles into documents. In Figure~\ref{fig:principle-evolution}, content revisions edit RD-03 and add and link related principles, and organization revisions edit an index entry, move a URL-parsing principle, and make a hub that groups redirects with URL parsing. A step applies primitives of only one kind, so that the effect of each kind can be measured on its own.

\begin{property}[Locality and traceability]
\label{prop:locality}
If $B \oplus \delta$ is defined, it differs from $B$ only in the targets of the primitives of $\delta$. Since every accepted revision is recorded and each primitive stamps its step in the provenance of what it changes, every library of a run is its seed followed by a recorded sequence of revisions, and each principle can be traced to its published source and to the revisions that changed it.
\end{property}


\subsection{Evolution as Constrained Search}
\label{sec:optimization}

Let $\mathcal{B}$ be the set of libraries reachable from the seed $B_0$ by revisions over $\mathcal{O}$. Evolution searches $\mathcal{B}$ for the $B^\star$ of Eq.~\ref{eq:objective}. Because $J$ and $P$ can only be estimated from executions, it proceeds as a local search with an acceptance criterion, one revision at a time,

\begin{equation}
B_{t+1} = \begin{cases} B_t \oplus \delta_t & \text{if } \delta_t \text{ meets acceptance criterion}, \\ B_t & \text{otherwise}, \end{cases}
\label{eq:step}
\end{equation}
where $\delta_t$ is proposed from the feedback on $B_t$, and the acceptance criterion (Definition~\ref{def:acceptance}) decides on fresh executions whether $B_t \oplus \delta_t$ improves on $B_t$ without leaving $\mathcal{F}$. Eq.~\ref{eq:step} is the learning procedure $L$ of \S\ref{sec:problem-formulation} for \ProjectName{}.

\lead{Feedback.} \ProjectName{} condenses each trajectory $\tau_B$ into a record $e = (R, W, \mathrm{diff}, f_x, s_x, Q)$, the sequence $R$ of library documents the agent opened, each marked as read before or after its first edit, the files $W$ it read and changed, its final diff, its outcome, and the functional and security test cases $Q$ that failed. The feedback $E(B)$ on a library is the set of records over all training tasks, and $E(B_t)$ is the $E_t$ of \S\ref{sec:problem-formulation}.

\lead{The loop.} A \emph{run} starts from the seed $B_0$ and repeats the evolution \emph{step} of Eq.~\ref{eq:step} until a stopping criterion holds (Figure~\ref{fig:overview}). Each step starts from the anchor $B = B_t$, the current library. 
\textsc{Test} executes the agent on every training task to produce $E(B)$. \textsc{Analyze}, \textsc{Draft}, and \textsc{Decide} then attribute failures (\S\ref{sec:diagnose}), draft a revision (\S\ref{sec:draft}), and estimate its effect (\S\ref{sec:decide}).
If the revision is accepted, the revised library becomes the new anchor, and its executions provide the feedback for the next step. Otherwise, the anchor stays unchanged, and the ledger keeps the rejected revision and its measurements for a later retry. Algorithm~\ref{alg:approach} in Appendix~\ref{app:impl} gives the pseudocode. 

Agent instances separate from the coding agent carry out the steps that need judgment, using the same model as the coding agent, as in ACE and SelfOp~\cite{zhang2026agentic,ullah2026selfop}, so no stronger model guides the evolution. Nothing from them reaches the coding agent except the library. Code applies the primitives and checks the acceptance criterion, so these instances may reject a revision that meets the criterion but cannot accept one that does not.

\lead{Stopping criterion.} A run ends after ten rejected candidates in a row, after three steps in a row with no candidate to evaluate, or at a step limit. A step has no candidate when no drafted revision survives the checks in \S\ref{sec:draft}, when its primitives do not apply to the library (Definition~\ref{def:revision}), or when the surviving revisions leave the index and every document's text unchanged.

\lead{Selection criterion.}
\label{sec:final}
Every kept library met the acceptance criterion against its predecessor, so the latest, $B_{\mathrm{last}}$, carries the most evidence and is the default. Training scores reflect the tasks that shaped each library, so the \emph{selection criterion} compares the kept libraries on validation tasks $\Tva$ and replaces $B_{\mathrm{last}}$ with the best of them, $B_{\mathrm{best}}$, only if the 95\% bootstrap interval of their paired difference lies above zero (Appendix~\ref{app:impl}). Because $\Tva$ shaped no revision, this choice also guards generalization over the course of a run, and the test tasks play no part in it. 

\subsection{Attributing Failures}
\label{sec:diagnose}

\begin{problem}[Attribution]
\label{prob:attribution}
Given the feedback $E(B)$ on a library $B$, assign each failed record a cause and the part of the library that failed, so that a revision changes that part.
\end{problem}


To guide revision, we distinguish four causes of failure $\mathcal{C}= \{\text{missing}, \text{unread}, \text{misapplied}, \text{too strict}\}$. Table~\ref{tab:causes} summarizes the evidence used to diagnose each cause and the mutation primitives that can address it.

Each cause $\gamma \in \mathcal{C}$ leaves its own evidence in the records and calls for primitives $\mathcal{O}_\gamma \subseteq \mathcal{O}$ that change the part that failed, so attributing a failure also determines which primitives a revision should use. In Figure~\ref{fig:principle-evolution}, the agent working with $B_t$ read RD-03, but its check let the protocol-relative target \texttt{//evil.com} through, a misapplied principle whose remedy is an edit of $k$.


\begin{table}[t]
\centering
\caption{Causes of failure at a situation $\sigma$.}
\label{tab:causes}
\scriptsize
\setlength{\tabcolsep}{3pt}
\begin{tabularx}{\columnwidth}{@{}>{\raggedright\arraybackslash}p{1.15cm}>{\raggedright\arraybackslash}p{2.2cm}L>{\raggedright\arraybackslash}p{1.6cm}@{}}
\toprule
Cause $\gamma$ & Condition at $\sigma$ & Evidence in the records & Primitives $\mathcal{O}_\gamma$ \\
\midrule
Missing & no $\pi$ has $a(\sigma)$ & security test fails & $\mathrm{add}$, $\mathrm{edit}(a)$ \\
Unread & $\pi$ with $a(\sigma)$ unopened & document not opened before the first edit & $\mathrm{entry}$, $\mathrm{move}$, $\mathrm{hub}$ \\
Misapplied & $\pi$ read, $k$ misses the property & security test fails after $\pi$ is read & $\mathrm{edit}(k)$, $\mathrm{edit}(a)$ \\
Too strict & $\pi$ enforced where $v$ holds & functional test fails on valid input & $\mathrm{edit}(v)$, $\mathrm{edit}(a)$ \\
\bottomrule
\end{tabularx}
\end{table}

\lead{\textsc{Analyze}.} \textsc{Analyze} solves Problem~\ref{prob:attribution} by turning $E(B)$ into findings $\Phi$. A finding $\varphi = (\gamma, \eta, X_\varphi)$ names a cause $\gamma \in \mathcal{C}$, gives a diagnosis $\eta$, and cites the executions $X_\varphi \subseteq E(B)$ that show it; their tasks are its targets $T_\varphi$. Separate analyses examine insecure working code, broken code, and which documents the agent opened, since each cause shows in different evidence. \textsc{Analyze} also \emph{probes} failing tasks, executing the agent again with the guidance ladder of \S\ref{sec:motivation}, up to the reference fix where the benchmark provides one, within 100 executions per library. A finding must cite real executions, and findings are ranked by coverage, $|T_\varphi|/|\Ttr|$, so revisions go where the agent fails most. After a rejection, \textsc{Analyze} also examines what the rejected candidate broke. The analyses thus see each failing test case's inputs and expected result, which can reveal the attack, but not the test code that checks them.

\subsection{Generalizing Revisions}
\label{sec:draft}

\begin{problem}[Generalization]
\label{prob:generalization}
Given findings $\Phi$, find a revision $\delta$ whose principles have the scope of a class of code, so that a gain on $\Ttr$ carries to task families that training never sees.
\end{problem}


\lead{\textsc{Draft}.} \textsc{Draft} solves Problem~\ref{prob:generalization} by stating each revision's principles for a class of code. It fills $a$ with a situation that covers the cited failures without naming their tasks, $k$ with the check that must hold there, and $v$ with the behavior that must remain allowed, using the primitives $\mathcal{O}_\gamma$ of the findings' causes. A revision $\delta$ addresses a set of findings $\Phi_\delta \subseteq \Phi$, and its \emph{targets} are the tasks those findings cite, $T_\delta = \bigcup_{\varphi \in \Phi_\delta} T_\varphi$, on which \textsc{Decide} measures it (\S\ref{sec:decide}).
Besides new findings, \textsc{Draft} may retry or revise a revision from the ledger, or combine small revisions so that their targets add up. In Figure~\ref{fig:principle-evolution}, the revised RD-03 applies when a redirect target comes from the request, a situation that names no task, function, or field. A drafted revision enters the candidate library only if its primitives apply and its text says what the revision intended, is complete, and neither contradicts the rest of the library nor weakens its published source. A further check sends back for rewriting any revised principle that reuses a training task's names or details (Appendix~\ref{app:impl}), which removes the most common way a drafted principle fits one task (Appendix~\ref{app:step}).


\subsection{Estimating Effects}
\label{sec:decide}

\begin{problem}[Estimation]
\label{prob:estimation}
Given the anchor $B$ and a candidate $B^+ = B \oplus \delta$, decide from finitely many executions whether $J(B^+) > J(B)$ and $B^+ \in \mathcal{F}$.
\end{problem}

Three things make a single comparison unreliable. A revision's effect can be smaller than the agent's run-to-run variation~\cite{miller2024errorbars}, security tests alone cannot distinguish secure code from code that refuses its input~\cite{cweval}, and the executions that selected a revision's targets failed partly by chance. \textsc{Decide} solves Problem~\ref{prob:estimation} with three estimates and an acceptance criterion.

\lead{Paired comparison on fresh executions.} \textsc{Decide} executes the agent with the candidate on the training tasks and, on the revision's targets, executes it afresh with both the candidate and the anchor (Appendix~\ref{app:impl}), so that the estimate is free of the selection bias that reusing the executions which selected the targets would introduce. The two libraries are compared task by task~\cite{miller2024errorbars},
\begin{equation}
\widehat\Delta = \frac{1}{|T|}\sum_{x\in T} d_x, \qquad \mathrm{se} = \frac{\max\big(\hat\sigma_d,\ 0.42/\sqrt{n}\big)}{\sqrt{|T|}},
\label{eq:paired}
\end{equation}
where $T$ is a set of tasks, $d_x = u^+_x - u^-_x$ is the difference between the two libraries in $j_x$ or $f_x$ on task $x$, $\hat\sigma_d$ is the standard deviation of the differences, and the floor, with $n = 1$ at function level, keeps a few agreeing tasks from looking decisive (Appendix~\ref{app:stats}). Security is compared only on tasks where both libraries' code works, so that a security revision gains only by making working code secure.

\lead{Evidence across rounds.} \textsc{Decide} measures a candidate in a \emph{round}, fresh executions of the agent with the candidate and with the anchor on the training tasks. One round rarely settles whether a revision helps, so every revision stays in a ledger $\mathcal{H}$ with its measurements. The rounds $r$ that measured a revision $\delta$ on $T_\delta$, each with a gain $\widehat\Delta_r$ and standard error $\mathrm{se}_r$ from Eq.~\ref{eq:paired}, are pooled by inverse-variance weighting~\cite{cochran1954combination}, with $v_r = 1/\mathrm{se}_r^2$ and partial weight for rounds measured on earlier versions (Appendix~\ref{app:impl}),
\begin{equation}
\bar\Delta_\delta = \frac{\sum_r v_r\,\widehat\Delta_r}{\sum_r v_r}, \qquad
\mathrm{se}_\delta = \frac{1}{\sqrt{\sum_r v_r}}.
\label{eq:pool}
\end{equation}

\lead{Effect on all tasks.}A revision must not cost more elsewhere than it gains on its targets. Its effect on all training tasks is a stratified estimate~\cite{cochran1977sampling},
\begin{equation}
\widehat\Delta_{\mathrm{all}} = w\,\bar\Delta_\delta + (1-w)\,\widehat\Delta_{\Ttr\setminus T_\delta}, \qquad w = \frac{|T_\delta|}{|\Ttr|},
\label{eq:strat}
\end{equation}
so that a precise gain on a few targets cannot dominate it.

\begin{definition}[Acceptance criterion]
\label{def:acceptance}
An estimate $\widehat\Delta$ with standard error $\mathrm{se}$ is a \emph{gain} if $\widehat\Delta > \tfrac12\,\mathrm{se}$, \emph{clearly positive} if $\widehat\Delta - 1.645\,\mathrm{se} > 0$, and \emph{clearly negative} if $\widehat\Delta + 1.645\,\mathrm{se} < 0$, with 1.645 the 95th percentile of the standard normal. A revision is accepted if it meets one route and no rejection.
\begin{itemize}[leftmargin=1em, itemsep=1pt]
\item \emph{Routes.} $S_{|P}$ or $P$ gains and the other is not clearly negative, or the effect on $T_\delta$ is clearly positive, $\widehat\Delta_{\Ttr\setminus T_\delta}$ is not clearly negative, and no $\widehat\Delta_{\mathrm{all}}$ is negative. An organization revision may also be accepted if $J$ gains, $K$ falls without fewer executions opening the library, or a hub moves only unopened documents.
\item \emph{Rejections.} $\widehat\Delta_{\mathrm{all}}$ of $J$ or $\bar\Delta_\delta$ is clearly negative, an organization revision clearly lowers how many executions open the library, or $\widehat m + 1.645\,\mathrm{se}(\widehat m) < -\epsilon_F$ for $\widehat m = \widehat P - \widehat P_\varnothing$, a one-sided non-inferiority test~\cite{wellek2010testing} of the floor.
\end{itemize}
\end{definition} 

\begin{property}[No accumulated functional loss]
\label{prop:floor}
Every candidate is tested against $B_\varnothing$ rather than against its predecessor, so the functional losses of successive revisions are always measured cumulatively, and no accepted library falls clearly below the floor $P(B_\varnothing) - \epsilon_F$.
\end{property}


A gain needs only half a standard error, so a revision with no effect is sometimes accepted, which trades some false acceptances for faster progress. Such revisions can accumulate over a run, Property~\ref{prop:floor} bounds what they can cost functional code, and EQ2 measures the net effect on held-out tasks.

\lead{Retrying and retirement.} \textsc{Draft} weighs revisions in the ledger against new findings by an optimistic estimate of their effect on all training tasks, $(|T_\delta|/|\Ttr|)(\bar\Delta_\delta + \mathrm{se}_\delta)$, in the manner of upper-confidence-bound bandits~\cite{auer2002finite}, so that a promising but rarely measured revision is retried. The \emph{retirement criterion} removes a revision once $\bar\Delta_\delta + 1.645\,\mathrm{se}_\delta < 0$ or after three rounds without a clear gain.

\section{Evaluation}
\label{sec:evaluation}

We evaluate whether coding agents equipped with \ProjectName{} gain security self-improvement. We use function-level secure coding tasks from two established benchmarks: the standard setting in which prior work measures secure code generation~\cite{safecoder2024,fu2024constrained,cweval,nie2026secodeplt}, and a full coding agent solves each task in its own workspace. Own experience holds by construction, so EQ1 and EQ2 test the other three conditions of Definition~\ref{def:ssi}, EQ3 asks whether improvement continues over a run, and EQ4 and EQ5 ask what the agents learn and how far it carries. 
We further examine the limits of \ProjectName{} on repository-level tasks from SusVibes~\cite{susvibes2026}, a recent benchmark built from real Python projects (\S\ref{sec:repo}).

\begin{itemize}[leftmargin=1em, itemsep=1pt]
\item \textbf{EQ1 (Persistence).} Do lessons persist across sessions, so that a fixed vulnerability does not recur?
\item \textbf{EQ2 (Generalization).} On unseen task families, do agents write more secure code without losing functional code, compared with fixed guidance and with methods that also learn from experience?
\item \textbf{EQ3 (Continued learning).} Do agents keep improving over a run, and which parts of \ProjectName{} carry the gains?
\item \textbf{EQ4 (What is learned).} Which CWEs improve, what do the learned principles state, and where do they fail?
\item \textbf{EQ5 (Transfer).} Do the learned principles carry to other agents, models, and benchmarks?
\end{itemize}

\subsection{Experimental Setup}
\label{sec:setup}

\lead{Coding agents.} We study three coding agents, Codex, Claude Code, and Qwen Code, which differ in scaffold but run on the same open model, Qwen3.6-35B-A3B (Appendix~\ref{app:impl}), with a budget of 100 turns per execution. \ProjectName{} learns a separate library for each from the same seed. Its judgment steps (\S\ref{sec:optimization}) run in separate Codex instances on the same model, each in a clean, isolated workspace that shares no session, memory, or files with the coding agent. For transfer across models (EQ5), we also learn a library with Qwen Code on the open Mistral Medium 3.5.

\lead{Benchmarks and splits.} SeCodePLT~\cite{nie2026secodeplt} has 818 Python tasks over 17 CWEs, which we group by text similarity into 90 task families of about nine tasks each (\S\ref{sec:motivation-setup}). Before any run, we split them by family into $\Ttr$, $\Tva$, and $\Tte$ (507, 182, and 129 tasks), so that every CWE appears in each split and no family spans two. CWEval~\cite{cweval} has 119 tasks over 31 CWEs in five languages, which we split by problem into 73 training and 46 test tasks, with no test CWE among the training tasks. 

\lead{Evolution and selection.} A run is limited to 100 steps, or 20 for the ablations of EQ3. In a round, the agent executes every training task once with each library, and the revision's target tasks four times. The main runs ended after 71, 56, and 69 steps, with 16, 14, and 17 kept libraries: Codex's after ten rejected candidates in a row, the other two after three steps in a row with no candidate. Selection on the validation tasks chose the latest kept library in every run except Codex’s, where the library kept at step 52 scored 3.7 points above the one kept at step 58. Appendix~\ref{app:impl} reports the evolution and deployment costs.

\lead{Delivery.} The library's index is placed in the task's workspace as the agent's project instruction file, \texttt{AGENTS.md} for Codex and Qwen Code and \texttt{CLAUDE.md} for Claude Code, which each agent loads on its own, with the documents beside it (\S\ref{sec:guidebook}). No system prompt changes. The task prompt gains one fixed sentence, the same for every library, including the seed: ``You MUST open and read the guidebook documents relevant to your task before writing any code, and follow them.''

\lead{Baselines.} We compare \ProjectName{} with two groups of alternatives. 
The first keeps the agent's guidance fixed and consists of no guidance, a one-line security reminder~\cite{susvibes2026}, and the same seed library $B_0$. 
The second learns from the agent's experience, including ACE~\cite{zhang2026agentic}, which evolves a context, SkillOpt~\cite{yang2026skillopt}, which optimizes a skill document, and RuleEvolve~\cite{jiang2026ruleevolve}, which evolves coding rules. 
These methods were designed to improve task performance in general, so we replace each method's optimization target with secure-pass@1, the objective $J$ of Eq.~\ref{eq:objective}. All three learn from single model calls on $\Ttr$ with the same model as the coding agents, ACE and SkillOpt through their released implementations and RuleEvolve through our reimplementation, since its code is not released, and each agent receives what they learn as its project instructions. RuleEvolve starts from the published principles, ACE from an empty playbook, and SkillOpt from the empty skill of its base configuration.

\lead{Metrics.}
\label{sec:metrics}
We report secure-pass@1 ($J$) and pass@1 ($P$) of Eq.~\ref{eq:jpk}, and two shares of functional code that is secure, $S_{|P}$, the share of all functional executions that are secure (\S\ref{sec:problem-formulation}), and secure@1$_{\text{pass}}$~\cite{fu2024constrained}, the share of each task's functional executions that are secure, averaged over tasks, with a task without functional code counting as zero.  
A rise in $S_{|P}$ alone means the agent secures the same tasks more often, and a rise in secure@1$_{\text{pass}}$ alone means it secures more tasks.
Both shares also rise when insecure code stops working, which \ProjectName{} limits by optimizing $J$ and keeping $P$ above the floor of Definition~\ref{def:ssi}.
Tables abbreviate them as sp@1, p@1, $S_{|P}$, and s@1$_{\text{p}}$. 
EQ1 measures persistence with RR and RG (\S\ref{sec:motivation-metrics}), and cost is $K$, the input tokens per execution. Each condition is executed ten times per test task, a choice Appendix~\ref{app:k} explains. Conditions are compared by their paired difference on the same tasks. We call a difference significant when it is above or below 0 with 95\% confidence, that is, when its 95\% bootstrap interval, over tasks or, for the chains, over chains, lies entirely on one side of 0.

\subsection{Results}
\label{sec:results}

\lead{EQ1: With \ProjectName{}, lessons persist across sessions.}\label{sec:persist} We rerun the 14 chains of the motivating study (\S\ref{sec:motivation-setup}), drawn from task families held out from training, with Codex, Claude Code, and Qwen Code on Qwen3.6, each with the final library $B^\star$ of its main run (\S\ref{sec:setup}). A \emph{pair} is one chain run by one agent, resulting in 42 pairs. We give the library in two ways. In (a), the agent receives it only at the later task, so the failure and fix at $T_i$ match those of the motivating study and its conditions compare directly. In (b), the agent has it at every task, as in deployment.

With the library only at the later task (Table~\ref{tab:persist}a), the 36 pairs whose fix passed both kinds of tests introduce the same CWE in 11.5\% of executions, against 27.5\% with no earlier fix and 26.7\% after the fix alone. This retention gain of 16.0 points holds at every distance and exceeds those of the same session (8.4) and the agent's own note (12.5). Recurrence is also lower than with any general guidance in the motivating study, including the principle restated by hand (16.7\%, level (d)), and close to guidance written for each task (levels (e) and (f), Table~\ref{tab:ladder_result}).

With the library at every task (Table~\ref{tab:persist}b), the agent wrote functional and secure code in all five executions at $T_i$ in 48\% of the pairs (20 of 42), against 7\% expected from each agent's own success rate, so the library often prevented the vulnerability before any fix. Of the other 22 pairs, 17 still wrote the vulnerability, fixed it, and have a chain in the motivating study to compare with, and the recurrence on these is 24.5\% with the library, against 32.0\% without it. 


\begin{table}[t]
\centering
\caption{Recurrence at the later task (\%).}
\label{tab:persist}
\scriptsize
\resizebox{\columnwidth}{!}{
\begin{tabular}{@{}lccccc@{}}
\toprule
& \multicolumn{3}{c}{RG$\uparrow$ by distance $m$} & \multicolumn{2}{c}{All $m$} \\
\cmidrule(lr){2-4}\cmidrule(l){5-6}
Condition & 0 & 3 & 6 & RR$\downarrow$ & RG$\uparrow$ \\
\midrule
\multicolumn{6}{@{}l}{\emph{(a) At the later task: 36 pairs, BIR 27.5}} \\
New session & $-1.1$ & $+0.0$ & $+3.3$ & 26.7 & $+0.7$ \\
Same session & $+10.6$ & $+7.1$ & $+7.4$ & 19.2 & $+8.4$ \\
Own note (new session) & $+13.6^{*}$ & $+12.5^{*}$ & $+11.4^{*}$ & 15.0 & $+12.5^{*}$ \\
\textbf{\ProjectName{} (new session)} & $+15.3^{*}$ & $+17.2^{*}$ & $+15.6^{*}$ & \textbf{11.5} & $+16.0^{*}$ \\
\midrule
\multicolumn{6}{@{}l}{\emph{(b) At every task: 17 pairs, BIR 32.4}} \\
New session & $-2.9$ & $+0.6$ & $+3.5$ & 32.0 & $+0.4$ \\
\textbf{\ProjectName{} (new session)} & $+11.2^{*}$ & $+4.7$ & $+7.6$ & \textbf{24.5} & $+7.8^{*}$ \\
\bottomrule
\end{tabular}
}
\par\smallskip\raggedright
$^{*}$: gain whose 95\% bootstrap interval over chains lies above 0.\par
\end{table}

\begin{table}[t]
\centering
\caption{SeCodePLT test tasks (\%).}
\label{tab:secodeplt}
\resizebox{\columnwidth}{!}{%
\begin{tabular}{llcccc}
\toprule
Agent & Guidance & p@1$\uparrow$ & sp@1$\uparrow$ & s@1$_{\text{p}}$$\uparrow$ & $S_{|P}$$\uparrow$ \\
\midrule
\multirow{7}{*}{Codex}
 & No guidance & 51.3 & 41.7 & 54.4 & 81.3 \\
 & Security reminder & 49.5 & 41.9 & 56.5 & 84.5 \\
 & Seed library & 50.7 & 42.5 & 57.8 & 83.8 \\
 & ACE & 44.2 & 37.5 & 47.3 & 84.9 \\
 & SkillOpt & 49.3 & 38.8 & 48.4 & 78.8 \\
 & RuleEvolve & \textbf{51.8} & 43.7 & 53.6 & 84.4 \\
 & \textbf{\ProjectName{}} & 51.4 & $\textbf{50.6}^{*}$ & \textbf{65.7} & \textbf{98.5} \\
\midrule
\multirow{7}{*}{Claude Code}
 & No guidance & 46.3 & 36.9 & 47.2 & 79.7 \\
 & Security reminder & 48.1 & 39.1 & 52.9 & 81.3 \\
 & Seed library & 48.9 & 41.5 & 53.2 & 84.8 \\
 & ACE & 42.1 & 36.9 & 53.9 & 87.7 \\
 & SkillOpt & 45.1 & 37.6 & 51.8 & 83.3 \\
 & RuleEvolve & 46.4 & 39.1 & 51.5 & 84.4 \\
& \textbf{\ProjectName{}} & \textbf{49.8} & $\textbf{48.5}^{*}$ & \textbf{62.7} & \textbf{97.1} \\
 
\midrule
\multirow{7}{*}{Qwen Code}
 & No guidance & 45.0 & 36.9 & 46.6 & 81.9 \\
 & Security reminder & 46.0 & 39.0 & 49.9 & 84.7 \\
 & Seed library & 47.6 & 40.7 & 54.7 & 85.5 \\
 & ACE & 40.3 & 35.0 & 46.9 & 86.7 \\
 & SkillOpt & 42.9 & 35.3 & 48.0 & 82.3 \\
 & RuleEvolve & 45.8 & 38.1 & 49.0 & 83.1 \\
 & \textbf{\ProjectName{} }& \textbf{48.9} & $\textbf{47.8}^{*}$ & \textbf{61.4} & \textbf{97.8} \\
\bottomrule
\end{tabular}%
}
\par\smallskip\raggedright\footnotesize
\textbf{Bold}: best per agent.$^{*}$: significant gain over no guidance.
\end{table}

\begin{figure}[t]
\centering
\includegraphics[width=\columnwidth]{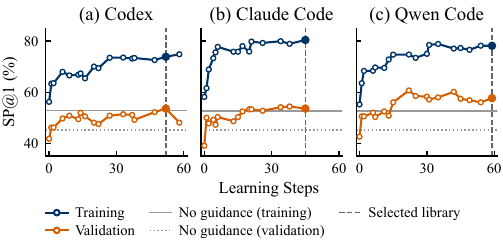}
\vspace{-3mm}
\caption{Secure-pass@1 of each library a run kept, at the step that kept it. The runs ended at steps 71, 56, and 69.}
\label{fig:curves}
\end{figure}

\lead{EQ2: On unseen task families, agents write more secure code without losing functional code.} On SeCodePLT's test tasks (Table~\ref{tab:secodeplt}), the security reminder and the seed library raise secure-pass@1 by at most 2.2 and 4.6 points, and neither gives any agent a significant gain. With the libraries they evolved from that seed, all three agents gain significantly, reaching 47.8--50.6\% secure-pass@1. 

The evolved libraries significantly improve security while preserving functionality. With them, 97.1--98.5\% of the agents' functional executions are secure ($S_{|P}$), against 80--82\% without guidance and at most 87.7\% under any baseline. This rise comes from insecure code becoming secure, since pass@1 does not fall (51.4, 49.8, and 48.9\% against 51.3, 46.3, and 45.0\% without guidance) and, on the same tasks, secure@1$_{\text{pass}}$ rises from 46.6--54.4\% to 61.4--65.7\%, the highest for every agent.

For every agent, \ProjectName{} also reaches higher secure-pass@1 than the methods that learn from the agents' experience with the same objective, model, and training tasks. ACE and SkillOpt perform close to no guidance, and RuleEvolve starts from the same published principles and ends close to the seed library. These results indicate that \ProjectName{}'s design is better suited to security self-improvement.

\lead{EQ3: Security improves as evolution proceeds, and ablations show what each component contributes.} Figure~\ref{fig:curves} plots the secure-pass@1 of every library each run kept, on the training tasks and on the validation tasks. A run keeps a new library about every four steps, and secure-pass@1 on the training tasks rises across the run, from 55--58\% to 74--81\%. On the validation tasks, it rises from 39--43\% to 54--58\%, more slowly and with more variation, and a quarter to two thirds of each run's gain there comes after its first kept library. The test tasks show the same pattern, as libraries kept midway through a run score below $B^\star$ (42.9 against 50.6\% for Codex and 45.3 against 47.8\% for Qwen Code). Learning continues because evidence accumulates across rounds, and 29 of the 96 revisions kept in these runs were kept only after more than one round.

Each ablation is a single 20-step run with Qwen Code that removes one component, compared with the full method's library at step 20 (Table~\ref{tab:ablation}), so differences of a few points are within the variation of one comparison (\S\ref{sec:decide}). Without the execution records, which show what the agent read, changed, and failed, or the revision ledger, which pools evidence across rounds, the run gains nothing over the seed, and without the analyses it gains little. Without the acceptance criterion, every applicable revision is kept, and the secure share of functional code falls from 96.4\% to 90.0\%.  Without the organization primitives, the result stays close to the full method's, since by step 20 the full run had used them only once, in a rejected revision. 
Over the full runs, 11 of the 47 kept libraries came from organization revisions, and one in Claude Code's run cut the input tokens per execution by 14\%.


\begin{table}[t]
\centering
\caption{Component ablations (\%).}
\label{tab:ablation}
\scriptsize
\resizebox{\columnwidth}{!}{%
\begin{tabular}{lccccc}
\toprule
Run & Kept & p@1$\uparrow$ & sp@1$\uparrow$ & s@1$_{\text{p}}$$\uparrow$ & $S_{|P}$$\uparrow$ \\
\midrule
\ProjectName{} & 7 & 47.4 & \textbf{45.7} & 56.2 & \textbf{96.4} \\
w/o organization primitives & 6 & 47.6 & 44.5 & \textbf{60.1} & 93.5 \\
w/o acceptance criterion & 19 & \textbf{48.1} & 43.3 & 56.1 & 90.0 \\
w/o analyses & 7 & 46.0 & 41.9 & 57.9 & 91.2 \\
w/o revision ledger & 4 & 45.9 & 40.7 & 55.0 & 88.8 \\
w/o execution records & 5 & 46.5 & 39.7 & 54.7 & 85.3 \\
\midrule
No guidance & -- & 45.0 & 36.9 & 46.6 & 81.9 \\
Seed library & -- & 47.6 & 40.7 & 54.7 & 85.5 \\
\bottomrule
\end{tabular}%
}
\end{table}

\lead{EQ4: Agents learn where and how to apply published principles.} By CWE (Appendix~\ref{app:cwe}), Codex's library helps most where a published principle lacked the situation or check that the agent's failures needed. On CWE-770 (allocation of resources without limits), for example, Codex never wrote secure code without guidance or with the seed library, and does so in 68.0\% of executions once its library gave the published principle the situation $a$ and the check $k$ it lacked. Training volume does not explain which CWEs gain or lose, since every CWE had 20 to 40 training tasks, and most losses are small or rest on a single test task, where one execution moves the rate by 10 points.

The largest loss, on CWE-77 (command injection), indicates that a learned principle can overreach. A principle learned from CWE-200 (exposure of sensitive information) tells the agent to fill a template's placeholders with Python's \texttt{str.format}, and its situation, a template string with placeholders, also matches the CWE-77 tasks, which fill HTML templates whose placeholders are written \texttt{\{\{name\}\}}. Since \texttt{str.format} reads \texttt{\{\{} as an escaped brace, Codex rendered \texttt{<p>Hello, \{\{name\}\}!</p>} as \texttt{<p>Hello, \{Alice\}!</p>}. Codex's code thus stays secure whenever it works, but less of it works, and executions that opened this principle failed their functional tests in 72\% of cases (33\% for the others).

\begin{table}[t]
\centering
\caption{Transfer (p@1\,/\,$S_{|P}$, \%).}
\label{tab:transfer}
\scriptsize
\setlength{\tabcolsep}{2.1pt}
\textbf{(a) Across agents}\\[2pt]
\begin{tabular}{lccccc}
\toprule
& & & \multicolumn{3}{c}{Learned with} \\
\cmidrule(l){4-6}
Evaluated with & None & Seed & Codex & Claude Code & Qwen Code \\
\midrule
Codex & 51.3\,/\,81.3 & 50.7\,/\,83.8 & \textbf{51.4}\,/\,\textbf{98.5}$^{*}$ & 49.5\,/\,96.9$^{*}$ & 50.1\,/\,98.1$^{*}$ \\
Claude Code & 46.3\,/\,79.7 & 48.9\,/\,84.8$^{*}$ & 46.8\,/\,98.0$^{*}$ & \textbf{49.8}\,/\,97.1$^{*}$ & 48.7\,/\,\textbf{98.2}$^{*}$ \\
Qwen Code & 45.0\,/\,81.9 & 47.6\,/\,85.5 & 48.0\,/\,\textbf{97.9}$^{*}$ & \textbf{51.2}\,/\,96.5$^{*}$ & 48.9\,/\,97.8$^{*}$ \\
\bottomrule
\end{tabular}
\\[6pt]\textbf{(b) Across models}\\[2pt]
\begin{tabular}{lcccc}
\toprule
& & & \multicolumn{2}{c}{Learned with} \\
\cmidrule(l){4-5}
Evaluated with & None & Seed & Qwen3.6 & Mistral \\
\midrule
Qwen3.6 & 45.0\,/\,81.9 & \textbf{47.6}\,/\,85.5 & 47.4\,/\,\textbf{96.4}$^{*}$ & 46.0\,/\,86.5$^{*}$ \\
Mistral & 53.2\,/\,81.9 & \textbf{55.0}\,/\,85.8$^{*}$ & 51.0\,/\,\textbf{97.3}$^{*}$ & 52.9\,/\,91.2$^{*}$ \\
\bottomrule
\end{tabular}
\\[6pt]\textbf{(c) Across benchmarks}\\[2pt]
\begin{tabular}{lccccc}
\toprule
& & \multicolumn{2}{c}{Learned on SeCodePLT} & \multicolumn{2}{c}{Learned on CWEval} \\
\cmidrule(lr){3-4}\cmidrule(l){5-6}
Evaluated on & None & Seed & Learned & Seed & Learned \\
\midrule
SeCodePLT (129) & 45.0\,/\,81.9 & 47.6\,/\,85.5 & 48.9\,/\,\textbf{97.8}$^{*}$ & 47.6\,/\,84.9 & \textbf{49.4}\,/\,85.1 \\
\midrule
CWEval, C (12) & \textbf{81.7}\,/\,70.4 & 75.0\,/\,82.2 & 70.0\,/\,72.6 & 79.2\,/\,\textbf{86.3} & 78.3\,/\,80.9 \\
\quad C++ (8) & 80.0\,/\,71.9 & 80.0\,/\,92.2 & 70.0\,/\,85.7 & \textbf{81.2}\,/\,89.2 & \textbf{81.2}\,/\,\textbf{92.3} \\
\quad Go (7) & 81.4\,/\,64.9 & 75.7\,/\,86.8$^{*}$ & \textbf{95.0}\,/\,91.2$^{*}$ & 81.4\,/\,\textbf{93.0} & 65.7\,/\,91.3 \\
\quad JavaScript (9) & \textbf{86.7}\,/\,60.3 & 73.3\,/\,84.8 & 60.0\,/\,77.8 & 74.4\,/\,85.1$^{*}$ & 71.1\,/\,\textbf{93.7}$^{*}$ \\
\quad Python (10) & 84.0\,/\,70.2 & 82.7\,/\,87.9 & \textbf{87.9}\,/\,88.6$^{*}$ & 79.0\,/\,89.9$^{*}$ & 85.0\,/\,\textbf{90.6}$^{*}$ \\
\quad All (46) & \textbf{82.8}\,/\,67.7 & 77.3\,/\,86.5$^{*}$ & 75.3\,/\,82.9$^{*}$ & 78.9\,/\,88.4$^{*}$ & 77.0\,/\,\textbf{89.0}$^{*}$ \\
\bottomrule
\end{tabular}
\par\smallskip\raggedright
Bold: best per row. $^{*}$: gain over no guidance whose 95\% bootstrap interval over tasks lies above 0.\par
\end{table}

\lead{EQ5: Learned principles keep functional code secure across agents and models, and on other benchmarks in the language they were learned from.}

\lead{Across agents.} Each agent can use another's library. With it, 96.5--98.2\% of the agent's functional code is secure, against 79.7--81.9\% without guidance, a significant rise in every case, about as large as with its own library, while pass@1 stays within the floor ($\epsilon_F=3$ points, Table~\ref{tab:transfer}a). What a run learns is not tied to the agent that learned it.



\lead{Across models.} We cap the Mistral run at 17 steps, so Table~\ref{tab:transfer}b compares the libraries each model had kept by step 17. In both directions, a library learned on one model significantly raises the share of the other model's functional code that is secure, while pass@1 stays within the floor, from 81.9 to 97.3\% on Mistral and from 81.9 to 86.5\% on Qwen3.6. The library learned on Qwen3.6 makes more of Mistral's functional code secure than Mistral's own (97.3 against 91.2\%), whereas after 17 steps, the library learned with Mistral helps Qwen3.6 about as much as the seed library (86.5 against 85.5\%), whose rise is not significant.

\lead{Across benchmarks.} We evaluate Qwen Code's SeCodePLT library on CWEval, whose tasks span five languages (Table~\ref{tab:transfer}c). On CWEval's Python tasks, it significantly raises the share of functional code that is secure (88.6 against 70.2\%) and keeps pass@1 above no guidance (87.9 against 84.0\%). Every library, the seed included, makes more of CWEval's functional code secure but lowers pass@1 beyond the floor, most for the library learned on SeCodePLT in C, C++, and JavaScript, languages absent from its training tasks. A library learned on CWEval's own 73 training tasks stays close to its seed on both benchmarks, since they cover 18 CWEs, mostly one problem in five languages, against 20--40 Python tasks per CWE in SeCodePLT, and include none of the test CWEs. Learning thus transfers within the language it was learned from and needs enough failures in the CWEs where it will be used.

\section{Toward Self-Improvement in Repositories}
\label{sec:repo}


Coding agents increasingly implement features inside existing projects. Since such tasks can also come with the functional and security tests that learning requires (\S\ref{sec:problem-formulation}), we apply \ProjectName{} to them, by evolving a library on them and by transferring the one learned on SeCodePLT, to mark where its current design applies.

\lead{Setting.} We use 184 Python feature tasks from SusVibes~\cite{susvibes2026}, drawn from 101 projects, each checked by the project's own tests and by the tests of a real security fix. We split them by project into 88 training, 48 validation, and 48 test tasks, evolve a library with Codex on Qwen3.6 for 20 steps, and run each test task three times.

\lead{Results.} No library makes repository code significantly more secure than no guidance (Table~\ref{tab:repo}). Under every condition, Codex writes functional code in about half of the executions, but only 10.5--13.5\% of it is secure, against 81.3--98.5\% on SeCodePLT. The few secure executions come from tasks Codex solves securely under every condition.

\lead{Challenges.} This experiment detects only large differences, but it points to three challenges in repositories, which we examine on the evolution executions and on the 30 test tasks where the evolved library left functional but insecure code, and \S\ref{sec:disc-repo} outlines how we will address them.

 \emph{(1) Securing repository code requires knowing the project.} A SeCodePLT task can be secured by following a principle for a class of code, such as resolving a path inside an allowed directory. About half of the 30 SusVibes tasks also require knowing how the project enforces security, such as the Vyper compiler's integer bounds or the Wagtail CMS's permission rules. Even where a principle applies, the fix can need a check specific to the project's code. In the Jinja template engine (CVE-2024-22195), Codex escapes attribute values as the output-encoding principle requires, but the vulnerability is an attribute key containing a space, which lets an attacker inject another attribute, such as an event handler.

 \emph{(2) Project-specific lessons are hard to learn.} During evolution, 98\% of the security findings concern a single task, against 70\% on SeCodePLT. A revision that secures only one of the 88 training tasks raises secure-pass@1 by at most about one point, which is small against the variation of a few executions. The run kept 4 libraries in 20 steps, each at or slightly above the seed when kept (10.3 to 11.9\% sp@1 on the training tasks, against 10.3\%), but measured again in later rounds they scored 7.6 to 10.2\%, so it never rose beyond the variation between rounds, whereas on SeCodePLT it rose from 55--58\% to 74--81\%. Since the split is by project, a lesson about a training project could not help on the test tasks anyway. 

 \emph{(3) Guidance does not reach the edits.} During evolution, a SusVibes execution makes a median of 108 tool calls, against 9 on SeCodePLT, and the library makes up only 0.11\% of its input tokens, against 5.6\%. On the 30 tasks, Codex never opened a library document on 10, and on the other 20 it returned to one after its first edit on only 4, so most of its edits come long after it last read a principle.

\begin{table}[t]
\centering
\caption{SusVibes test tasks with Codex on Qwen3.6 (\%).}
\label{tab:repo}
\begin{tabular}{@{}lcccc@{}}
\toprule
Guidance & p@1$\uparrow$ & sp@1$\uparrow$ & s@1$_{\text{p}}$$\uparrow$ & $S_{|P}$$\uparrow$ \\
\midrule
No guidance & 51.4 & 6.9 & 8.7 & 13.5 \\
Seed library & 52.8 & 5.6 & 5.9 & 10.5 \\
Learned on SeCodePLT & 52.1 & 6.9 & 7.3 & 13.3 \\
Learned on SusVibes & 53.5 & 6.9 & 9.7 & 13.0 \\
\bottomrule
\end{tabular}
\end{table}

\section{Related Work}
\label{sec:related}

\lead{Secure code generation.} Studies and benchmarks have measured the security of code written by language models and agents, from single functions~\cite{pearce2022asleep,siddiq2022securityeval,tony2023llmseceval,bhatt2023cyberseceval,cweval,nie2026secodeplt} to applications~\cite{baxbench} and real repositories~\cite{secrepobench2025,realsecbench2026,susvibes2026,securevibebench2026}, and have found that functionally correct code can be vulnerable even when no adversary is involved~\cite{susvibes2026,peng2025fcv}. They measure security one task at a time, whereas our motivating study follows a vulnerability after it has been fixed and shows that it recurs. Prior work makes generated code more secure by hardening the model through training, steering, or constrained decoding~\cite{sven2023,mixturecorrections2025,safecoder2024,secureinstruct2025,zhang2024seccoder,hajipour2024hexacoder,hasan2025teaching,xu2024prosec,seccoderx2026,liu2026purpcode,secpi2026,fu2024constrained}, or by guiding generation at inference time through prompt optimization, self-critique, analysis loops, or retrieved security guidance~\cite{promsec2024,datta2025reflexion,nunez2024autosafecoder,patir2025grasp,saul2025scgagent,mamun2026secgent,lin2025codeguarder}. Their security knowledge, however, is fixed before deployment and does not change when the agent writes vulnerable code. We instead study security self-improvement (Definition~\ref{def:ssi}), in which a coding agent learns from its own failures, and \ProjectName{} gives coding agents this capability.


\lead{Self-improving agents.} Agents improve from experience through verbal reflection~\cite{shinn2023reflexion}, self-critique~\cite{madaan2023selfrefine,kim2023rci}, skills~\cite{wang2023voyager}, induced rules~\cite{zhao2024expel}, reusable workflows~\cite{wang2025awm}, and scaffolding code~\cite{zelikman2024stop}. In work on recursive self-improvement, coding agents rewrite their own code~\cite{robeyns2025sica,zhang2026dgm}, and other work evolves the guidance they read, as prompts~\cite{agrawal2026gepa}, skills~\cite{yang2026skillopt}, playbooks~\cite{zhang2026agentic}, coding rules~\cite{jiang2026ruleevolve}, or specifications~\cite{cheng2026specforge}. ACE and SkillOpt also constrain how guidance changes, ACE through itemized, localized updates and SkillOpt by bounding each edit and accepting it only after a check on held-out tasks. These methods target task success or efficiency and learn from attempts that fail the task, whereas security self-improvement must learn from vulnerabilities in code that succeeds at its task and improve security without costing functionality. Inspired by methods that evolve an agent's guidance~\cite{zhang2026agentic,cheng2026specforge}, \ProjectName{} attributes each failure to the part of a principle that failed and accepts a revision only when fresh, paired executions show that it helps while functional code stays above a floor. Given the same objective, ACE, SkillOpt, and RuleEvolve do not reach its gains (\S\ref{sec:results}).

\lead{Self-improvement in security agents.} Self-improvement has also been applied to agents whose task is itself a security task. SelfOp~\cite{ullah2026selfop} optimizes a security agent's context, including its instructions, skills, and reference documents, much as \ProjectName{} evolves what a coding agent reads, but to reproduce known vulnerabilities, and Hu et al.~\cite{hu2026evorepair} accumulate repair experience to fix known vulnerabilities. Their objective is success at that task, which the task's outcome measures directly and knowledge of the known vulnerability supports. A coding agent's vulnerabilities instead arise while it succeeds at an unrelated functional task, with none known in advance and with functionality that must be preserved. Security self-improvement targets this setting, where their objectives have no counterpart, so we do not use them as baselines.

\lead{Agent defenses.} Defenses for agents either enforce policies over agent actions~\cite{chen2025shieldagent,wang2026agentspec,palumbo2026forge,shi2025progent,tsai2025conseca,debenedetti2025camel,jia2025taskshield,xiang2025guardagent} or evolve them across tasks, as SED distills harmful trajectories into retrieved policies~\cite{le2026sed}, AGrail maintains a lifelong memory of safety checks~\cite{luo2025agrail}, and recent work evolves the agent's harness~\cite{ruan2026hard,qu2026she,li2026evosafeharness,mao2026safeevolve}. These defenses address attacks and unsafe agent behavior. \ProjectName{} addresses vulnerabilities introduced during ordinary coding without adversarial manipulation, complementing defenses over the agent's actions (\S\ref{sec:disc-robust}).

\section{Discussion}
\label{sec:discussion}

\subsection{Repository-Level Security Self-Improvement}
\label{sec:disc-repo}

Our repository evaluation covers one benchmark, one agent, and a single short run (\S\ref{sec:repo}), so it can detect only large effects, and no library helped. We outline an action for each challenge it identified. The first two actions need tasks ordered in time within a project, which SusVibes, with fewer than two tasks per project, lacks, and the third changes the agent's scaffold, which \ProjectName{} keeps fixed.

\lead{Seeding project principles.} Securing repository code requires knowing the project (challenge (1)). We will seed a library for each project from the functions its security tests exercise and its earlier security fixes changed, such as validators and permission checks. Each function yields a principle whose situation names the code that must use it and whose check names the function, and a hub groups these principles under one index entry per project.

\lead{Learning within a project.} Project-specific lessons are hard to learn (challenge (2)). We will learn within each project by replaying its history as tasks, evolving the library on earlier tasks and evaluating it on later ones, where failures may recur in the same code. A principle whose situation names one project applies only to its code, so acceptance can rest on that project's tasks, which also measure persistence.

\lead{Delivering principles at each edit.} To test whether guidance fails to reach the edits (challenge (3)), we will place the relevant principles in the agent's context at each edit. An improvement would confirm this diagnosis, and no improvement would point to missing project knowledge. If confirmed, principles tied to a project's files can be retrieved when the agent edits those files or checked as policies by runtime enforcement~\cite{wang2026agentspec,palumbo2026forge}.

\subsection{Adversarial Robustness of \ProjectName{}}
\label{sec:disc-robust}

We consider four attacks on \ProjectName{}, what its design already prevents, and how the rest could be defended.

\lead{Principle poisoning.} A few consistent records suffice to poison evolvers that judge usefulness themselves~\cite{chen2026poisonedevolution}. In \ProjectName{}, poisoned records can at most lead to a draft, which fresh executions must still accept. Text planted in test inputs that the analyses read~\cite{liu2025aishell} could, however, attach to a useful revision a narrow exception $v$ that no test exercises. Requiring each allowed behavior to cite a failed functional test would block such exceptions.

\lead{Criterion evasion.} A contributed task family whose tests reward insecure code~\cite{roesner2026trusting} can make a harmful revision meet the acceptance criterion. Requiring gains on families from different contributors would stop a single contributor, but it would also slow the learning of principles that only one family needs.

\lead{Situation evasion.} A principle applies only where the agent judges that its situation matches, so an issue or code comment claiming that a URL is validated elsewhere could lead the agent to skip it. Situations that a static analyzer can express could instead be checked by runtime enforcement (\S\ref{sec:disc-repo}), without relying on the agent's judgment.

\lead{Conflict induction.} A principle whose situation is too broad can break code it does not fit, as the CWE-200 principle did on CWE-77 (EQ4). An adversary could induce such a loss on one CWE, which passes the rejections and the floor because they consider all training tasks together. Comparing executions that opened each principle with those that did not would expose such a principle from existing records.

Evaluating these defenses needs, for each attack on learning, a full evolution run with and without its defense, and tasks whose contributors are known, which our benchmarks do not record. We will build such tasks, starting with a task family whose tests reward insecure code, and measure how often its revisions are accepted. We will also study online learning, where anyone who interacts with the agent could shape what it learns.

\subsection{Limitations}
\label{sec:limitations}

\lead{Security is judged by tests.} Each benchmark's security tests check only the vulnerabilities its authors anticipated, so a library could learn to pass these tests without making code secure in general. Held-out task families and a second benchmark reduce this risk but do not remove it.

\lead{Most results come from one model and one run per agent.} Our main results use Qwen3.6 with three coding agents on single-function Python tasks, and each agent has a single evolution run, so we have not measured how much the learned library varies between runs. The three libraries, evolved in separate runs, give every agent a similar share of secure functional code (96.5--98.2\%, Table~\ref{tab:transfer}a), which suggests that the gains do not rest on one favorable run. Across benchmarks, a library helps only in the language it was learned from (EQ5), and \S\ref{sec:disc-repo} discusses repository tasks, where no library helped.

\lead{Evolution is costly.} Each main run executed the coding agent 35,600 to 46,400 times, with 0.31 to 0.39M input tokens per execution (Appendix~\ref{app:impl}). This cost is paid once per agent and model, and a library also serves other agents and models (EQ5), but at deployment, an execution with the final library reads 0.25 to 0.30M input tokens, compared with 0.11 to 0.16M without guidance.
\section{Conclusion}
\label{sec:conclusion}

Coding agents that fix a vulnerability often reintroduce the same type of vulnerability in later sessions, and neither their persistent context nor published secure coding principles reliably prevents this. We framed the missing capability as security self-improvement and built \ProjectName{}, which evolves a library of principles from an agent's own security failures and keeps a revision only when fresh, paired executions show that it helps while functional code stays above a floor. With it, recurrence in new sessions fell from 26.7\% to 11.5\%, and on unseen task families 97.1--98.5\% of the agents' functional code was secure while pass@1 held steady. The libraries also carried across agents and models, and because each is text, developers can read and audit what their agents have learned.

\bibliographystyle{IEEEtran}
\bibliography{ref}

\appendices

\section{A Step in Detail}
\label{app:step}

We follow one revision of Qwen Code's main run through the four phases and across three steps, from its draft in step 9
to its keep in step 12. Every number comes from the steps' own records.

\lead{\textsc{Test}.} The anchor is the run's fourth kept library, kept at step 7, with scored executions on all 507 training tasks. Code condensed
each execution into a record of the documents it opened, the files it changed, its final diff, and the test cases that failed.

\lead{\textsc{Analyze}.} The analyses wrote 85 findings: 46 on broken code, 25 on insecure working code, and 14 on which documents the agent opened and how the library groups them, and code ranked them by coverage. One finding on broken code named the cause \emph{too strict}. A principle added earlier in the run, \texttt{html-output-formatting/space-separator},
tells the agent to join rendered HTML elements with a space; agents applied it to list items inside \texttt{<ul>} and
\texttt{<ol>} containers as well, and the functional tests of four CWE-79 task families, which expect no separator there,
failed.

\lead{\textsc{Draft}.} \textsc{Draft} planned a content step of three revisions: a second round of a revision from
the ledger (a template-validation principle), a new principle for evaluating arithmetic expressions safely (W3), and W2,
a revision of the space-separator principle with an explicit exception for list items. W2's drafting call read the
finding and the principle, previewed its primitives, and edited the principle's statement $s$ and check $k$ so that the separator is required only outside list containers; code applied the two primitives, and the draft also gained a compliant list-item example.

\lead{\textsc{Decide}.} The agent executed all 507 training tasks with the candidate, and on each target task four times together with the anchor.
The acceptance criterion rejected it: functional code fell clearly over all tasks ($\Delta P = -3.6$ points, 95\% upper bound $-0.5$),
and W3 was clearly harmful on its own targets ($-41.7$ points of $J$) and reused a training task's variable name,
\texttt{user\_expr}, which the check against training-task details flagged. W2's measurements stayed in the
ledger.

\lead{The ledger carries the revision.} Ranked by its optimistic estimate, W2 came back in step 10, together with the template-validation revision. Its targets now
showed a clear gain pooled over two rounds, $+49.9$ points of $J$; on one task, \texttt{cwe-079\_\_969}, the candidate
wrote functional and secure code in five of five executions against the anchor's one of five. A revision counts only for
its share of the training tasks, however, and overall, out of 507, the estimate of $\Delta P$ was $-1.4$ points, short of a gain on $P$, and $S_{|P}$ barely moved ($+0.1$). It was rejected again and stayed in the
ledger. In step 12 it returned for a third round. The agent was executed afresh with the anchor, and its secure-pass@1 on the training tasks
measured 68.5, against 70.0 at step 8: the same library, differing by chance alone. W2's pooled gain on its targets
reached $+57.4$ points over three rounds; the candidate met the security route with a gain on $S_{|P}$, $+2.8$ points
against half a standard error of $+1.1$, no rejection applied (functional code stood 6.9 points above no guidance, far from the floor at $-3$), and it was kept, with a new principle, as the run's fifth library. A revision that one round could not judge was thus neither discarded nor kept on luck: the ledger
accumulated its evidence until the acceptance criterion could decide.

\lead{Checks in the main runs.} The check against training-task details fired on 147 of the 338 drafted revisions of the three main runs (43 to 44\% per run), most often for a task's own names and then for language-specific details, test fixtures, and weakened sources. Nine of them were dropped. The floor never rejected on its own a revision that a route would have accepted. In a calibration on recorded executions with an earlier version of the acceptance criterion, whose nominal false-acceptance rate was 10\%, a revision with no effect was accepted 68\% of the time when judged on the executions that selected its targets and 12\% when judged on fresh ones.

\section{The Motivating Study in Detail}
\label{app:motivation}

\lead{The chains.} Table~\ref{tab:chain-tasks} lists the 14 chains by SeCodePLT task number. Each agent fixes its own failed execution of $T_i$. The fix message lists the failing cases as the evaluator reports them (input, expected outcome, what the code did), names no weakness and suggests no fix; the note message asks the agent to ``write what you learned from this fix to [its instructions file], a file you read at the start of every future task''. The same session continues one conversation across the chain (Codex's \texttt{exec resume}, \texttt{claude --continue}, \texttt{qwen --continue}); a new session starts from the agent's home directory as the previous task left it.

\begin{table}[t]
\centering
\caption{The 14 chains (SeCodePLT task numbers).}
\label{tab:chain-tasks}
\scriptsize
\begin{tabular}{l c c c}
\toprule
CWE & $T_i$ & $T_j$, same family & $T_j$, other family \\
\midrule
CWE-77 & 869 & 865 & 851 \\
CWE-79 & 996 & 994 & 1001 \\
CWE-94 & 1128 & 1130 & 1377 \\
CWE-95 & 1180 & 1181 & 1211 \\
CWE-200 & 1266 & 140 & 1395 \\
CWE-327 & 1290 & 353 & 362 \\
CWE-327 & 360 & 361 & 354 \\
CWE-347 & 1298 & 426 & 440 \\
CWE-601 & 644 & 656 & 677 \\
CWE-770 & 1348 & 913 & 891 \\
CWE-770 & 896 & 898 & 913 \\
CWE-863 & 1216 & 1223 & 1249 \\
CWE-915 & 1370 & 1074 & 1393 \\
CWE-1333 & 59 & 58 & 1394 \\
\bottomrule
\end{tabular}
\end{table}

\begin{table}[t]
\centering
\caption{The ladder for one later task (CWE-770).}
\label{tab:ladder}
\scriptsize
\setlength{\tabcolsep}{3pt}
\begin{tabularx}{\columnwidth}{@{}cL@{}}
\toprule
Level & Sample guidance added to the task's prompt \\
\midrule
(a) & ``Make sure to follow best practices for security and avoid common vulnerabilities.'' \\
(b) & All 218 principles of the OWASP Secure Coding Practices~\cite{owasp_scp}, e.g., ``Validate data length.'' \\
(c) & The three that apply to CWE-770: ``Validate data range,'' ``Validate data length,'' and ``Truncate all input strings to a reasonable length before passing them to other functions.'' \\
(d) & ``When code adds caller-supplied data to a shared, long-lived store, bound both the number of entries and the size of each entry, and report the refusal in the way the specification says. Data within the bounds must be stored as before.'' \\
(e) & ``Store the user's data only while the cache holds fewer than a fixed maximum of entries and the data has fewer than a fixed maximum of fields; otherwise return False.'' \\
(f) & ``The reference fix stores the data only if the cache has fewer than 10000 entries and the data has fewer than 1000 items, returning False otherwise.'' \\
\bottomrule
\end{tabularx}
\end{table}

\lead{The guidance ladder.} Table~\ref{tab:ladder} presents the six levels of guidance of \S\ref{sec:motivation-setup} for one later task, on CWE-770.

\lead{Coverage of the later tasks.}
Two models rated, independently, whether one of the 218 OWASP principles directly addresses each later task's vulnerability, given the task and the full list: Claude, whose ratings give CC, and GPT (gpt-6.1-sol), which never saw them. A principle addresses a task directly when a developer who followed it as written would write the secure version, and partially when it concerns the same area but leaves the specific check to the developer. The two agree on 25 of the 27 later tasks (Cohen's $\kappa = 0.83$): GPT counts 17 as directly addressed against Claude's 19, and both disagreements are tasks Claude rated direct and GPT partial (a bound on a cache's entries for CWE-770, and an allow-list of updatable fields for CWE-915). With GPT's ratings, CC would be 63\% instead of 70\%.

\lead{An earlier one-step study.} The preliminary study of \S\ref{sec:motivation} (Table~\ref{tab:retention}) drew its 135 fixes from training and validation tasks, up to four per CWE and agent. In a new session, an agent that had fixed the vulnerability could not be told apart from one that never saw it, with differences in secure-pass@1 of $-1.7$ to $+3.4$ points on the same family and every 95\% interval spanning zero. In the same session, it wrote functional and secure code in 71 to 79\% of executions on the same family. The verbatim record recovered half of the same-session gain on the same family for Codex, a third for Claude Code, and two-thirds for Qwen Code. On another family, it raised secure-pass@1 for Codex and Claude Code by 9 to 10 points but lowered it for Qwen Code from 60.0\% to 50.9\%.

\begin{table}[t]
\centering
\caption{The earlier one-step study: secure-pass@1 (\%) on a later task of the same CWE, from the repaired task's
family or from another family.}
\label{tab:retention}
\footnotesize
\setlength{\tabcolsep}{1.6pt}
\begin{tabular*}{\columnwidth}{@{\extracolsep{\fill}}@{}lcccccc@{}}
\toprule
& \multicolumn{2}{c}{Codex $\uparrow$} & \multicolumn{2}{c}{Claude Code $\uparrow$} & \multicolumn{2}{c}{Qwen Code $\uparrow$} \\
\cmidrule(lr){2-3}\cmidrule(lr){4-5}\cmidrule(l){6-7}
Later task's family & same & other & same & other & same & other \\
\midrule
No earlier fix                         & 31.6 & 52.6 & 31.4 & 54.3 & 25.7 & \best{60.0} \\
Same session as fix & \best{71.6} & \best{67.4} & \best{70.9} & \best{70.9} & \best{79.4} & \best{60.0} \\
New session after fix & 32.6 & 60.5 & 29.7 & 52.6 & 29.1 & \second{54.3} \\
New session, verbatim record & \second{51.1} & \second{62.6} & \second{44.6} & \second{62.9} & \second{62.3} & 50.9 \\
\bottomrule
\end{tabular*}
\end{table}

\section{Results by CWE}
\label{app:cwe}

Table~\ref{tab:cwe} gives Codex's results on SeCodePLT's test tasks for each of the 17 CWEs, with the number of training tasks ($n_{\text{tr}}$) and test tasks ($n_{\text{te}}$) of each. Several CWEs have a single test task, so their rates rest on ten executions.

\begin{table*}[t]
\centering
\caption{Codex by CWE on SeCodePLT (\%).}
\label{tab:cwe}
\resizebox{\textwidth}{!}{%
\begin{tabular}{rcrrcccccccccccc}
\toprule
 & & & & \multicolumn{3}{c}{p@1$\uparrow$} & \multicolumn{3}{c}{sp@1$\uparrow$} & \multicolumn{3}{c}{s@1$_{\text{p}}$$\uparrow$} & \multicolumn{3}{c}{$S_{|P}$$\uparrow$} \\
\cmidrule(lr){5-7}\cmidrule(lr){8-10}\cmidrule(lr){11-13}\cmidrule(l){14-16}
CWE & Name & $n_{\text{tr}}$ & $n_{\text{te}}$ & None & Seed & \ProjectName{} & None & Seed & \ProjectName{} & None & Seed & \ProjectName{} & None & Seed & \ProjectName{} \\
\midrule
74 & Injection (downstream) & 40 & 10 & 4.0 & 9.0 & 6.0 & 2.0 & 3.0 & 1.0 & 15.0 & 30.0 & 3.3 & 50.0 & 33.3 & 16.7 \\
77 & Command injection & 21 & 10 & 69.0 & 62.0 & 46.0 & 69.0 & 62.0 & 46.0 & 100.0 & 100.0 & 100.0 & 100.0 & 100.0 & 100.0 \\
79 & Cross-site scripting & 40 & 1 & 100.0 & 100.0 & 90.0 & 100.0 & 100.0 & 80.0 & 100.0 & 100.0 & 88.9 & 100.0 & 100.0 & 88.9 \\
94 & Code injection & 40 & 1 & 100.0 & 70.0 & 80.0 & 100.0 & 70.0 & 80.0 & 100.0 & 100.0 & 100.0 & 100.0 & 100.0 & 100.0 \\
95 & Eval injection & 30 & 11 & 94.5 & 94.5 & 87.3 & 81.8 & 91.8 & 87.3 & 86.9 & 97.2 & 100.0 & 86.5 & 97.1 & 100.0 \\
200 & Information exposure & 40 & 1 & 80.0 & 70.0 & 90.0 & 70.0 & 70.0 & 90.0 & 87.5 & 100.0 & 100.0 & 87.5 & 100.0 & 100.0 \\
327 & Broken cryptography & 34 & 3 & 6.7 & 56.7 & 63.3 & 0.0 & 0.0 & 63.3 & 0.0 & 0.0 & 100.0 & 0.0 & 0.0 & 100.0 \\
347 & Signature verification & 31 & 10 & 23.0 & 27.0 & 28.0 & 22.0 & 27.0 & 28.0 & 68.3 & 80.0 & 50.0 & 95.7 & 100.0 & 100.0 \\
352 & Cross-site req.\ forgery & 20 & 10 & 42.0 & 34.0 & 56.0 & 42.0 & 34.0 & 56.0 & 100.0 & 90.0 & 100.0 & 100.0 & 100.0 & 100.0 \\
502 & Unsafe deserialization & 31 & 10 & 0.0 & 0.0 & 0.0 & 0.0 & 0.0 & 0.0 & 0.0 & 0.0 & 0.0 & -- & -- & -- \\
601 & Open redirect & 21 & 20 & 44.5 & 45.0 & 49.5 & 35.5 & 41.0 & 49.0 & 39.3 & 45.9 & 54.5 & 79.8 & 91.1 & 99.0 \\
770 & Unbounded allocation & 31 & 10 & 81.0 & 69.0 & 69.0 & 0.0 & 0.0 & 68.0 & 0.0 & 0.0 & 98.9 & 0.0 & 0.0 & 98.6 \\
862 & Missing authorization & 30 & 10 & 100.0 & 100.0 & 100.0 & 100.0 & 100.0 & 100.0 & 100.0 & 100.0 & 100.0 & 100.0 & 100.0 & 100.0 \\
863 & Incorrect authorization & 21 & 10 & 0.0 & 0.0 & 0.0 & 0.0 & 0.0 & 0.0 & 0.0 & 0.0 & 0.0 & -- & -- & -- \\
915 & Mass assignment & 20 & 1 & 100.0 & 100.0 & 100.0 & 90.0 & 90.0 & 80.0 & 90.0 & 90.0 & 80.0 & 90.0 & 90.0 & 80.0 \\
918 & Server-side req.\ forgery & 31 & 10 & 100.0 & 98.0 & 98.0 & 100.0 & 98.0 & 98.0 & 100.0 & 100.0 & 100.0 & 100.0 & 100.0 & 100.0 \\
1333 & Regular-expression DoS & 26 & 1 & 100.0 & 100.0 & 100.0 & 60.0 & 80.0 & 100.0 & 60.0 & 80.0 & 100.0 & 60.0 & 80.0 & 100.0 \\
\midrule
 & All & 507 & 129 & 51.3 & 50.7 & 51.4 & 41.7 & 42.5 & 50.6 & 54.4 & 57.8 & 65.7 & 81.3 & 83.8 & 98.5 \\
\bottomrule
\end{tabular}%
}
\par\smallskip\raggedright\footnotesize
--: no functional code.
\end{table*}

\section{Implementation Details}
\label{app:impl}

\begin{algorithm}[t]
\caption{Evolution of the principle library in \ProjectName{}.}
\label{alg:approach}
\scriptsize
\begin{algorithmic}[1]
\Require agent $\mathcal{A}$, seed $B_0$, training tasks $\Ttr$, validation tasks $\Tva$, tolerance $\epsilon_F$
\Ensure final library $B^\star$
\State $B \gets B_0$;\ $E \gets \textsc{Test}(B, \Ttr)$
\State ledger $\mathcal{H} \gets \varnothing$;\ kept $\mathcal{K} \gets (B_0)$
\Repeat
  \State $\Phi \gets \textsc{Analyze}(E, B)$ \Comment{attribution}
  \State $\delta \gets \textsc{Draft}(\Phi, \mathcal{H}, B)$ \Comment{generalization}
  \State $(\mathit{accept}, \mathcal{H}) \gets \textsc{Decide}(B, B \oplus \delta, \mathcal{H})$ \Comment{estimation}
  \If{$\mathit{accept}$}
    \State $B \gets B \oplus \delta$;\ $E \gets$ executions of $B$;\ append $B$ to $\mathcal{K}$
  \EndIf
\Until{the stopping criterion holds}
\State \Return $B^\star \gets \textsc{Select}(\mathcal{K}, \Tva)$ \Comment{\S\ref{sec:final}}
\end{algorithmic}
\end{algorithm}

Algorithm~\ref{alg:approach} summarizes the evolution procedure described in \S\ref{sec:optimization}. The remaining details specify its implementation and experimental settings.


\lead{Executions per task.}\label{app:k} Every function-level evaluation makes ten executions per task, and every SusVibes evaluation three. For SusVibes without guidance (Codex), the
running mean over the first $n$ executions, on the tasks completed to $n$, is $J = 7.7$ and $P = 52.6$ at $n = 10$ (all 184 tasks), and stays between 7.5 and 7.8 for $J$ and between 52.6 and 53.7 for $P$ up to $n = 17$: the uncertainty
barely shrinks beyond ten executions, because the spread between tasks dominates it. The function-level baselines have twenty executions per task, so there we can measure what $n = 10$ loses. On SeCodePLT's test tasks and on CWEval, every agent's secure-pass@1 and pass@1 at $n = 10$ lie within 1.0 point of their values at $n = 20$, and their standard errors are at most 5\% larger. A paired difference loses more, since the spread between tasks cancels in it: the security reminder's effect on Qwen Code's secure-pass@1 has a standard error of 1.00 points at $n = 10$ against 0.82 at $n = 20$ on SeCodePLT, and 2.44 against 2.23 on CWEval, while its estimate moves by 0.6 and 0.9 points.

\lead{Seed sources.} Python principles come from the OpenSSF guide~\cite{openssf_python}, the Python documentation~\cite{python_security}, and Bandit~\cite{bandit}, and, for CWEval's other languages, those for C and C++ from the SEI CERT standards~\cite{cert_coding}, those for Go from OWASP's Go Secure Coding Practices~\cite{owasp_go_scp}, the Go documentation~\cite{go_security}, and gosec~\cite{gosec}, and those for JavaScript from eslint-plugin-security~\cite{eslint_security}, Node.js Best Practices~\cite{nodebestpractices}, and OWASP's Node.js cheat sheets~\cite{owasp_cheatsheets}. An agent sees only the language-independent documents and those of its task's language, and on C++ tasks also those for C.

\lead{Infrastructure.} Every run serves the model with vLLM, with two GPUs per model server (four for Mistral Medium 3.5), and executes each coding agent in a CPU-only, network-isolated Apptainer container that reaches the model through a relay.

\lead{\textsc{Test}.} Executions whose infrastructure failed, such as a container that did not start or a scorer that timed out before a verdict, are executed again rather than scored, so that they cannot count against a library.

\lead{\textsc{Draft}.} A step's revisions target at most 24 tasks, shared among them. Revisions that change disjoint documents are drafted in parallel. A revised principle that reuses a training task's names or details is returned for rewriting up to twice and dropped if it still does, with the reason recorded in the ledger. In Eq.~\ref{eq:pool}, a revision of an earlier entry counts its parent's rounds at half weight, as does a round measured on an earlier anchor that changed the same documents. In one run, the check against the published source sent back a revision that named only MD5 as a forbidden hash, since its source also forbids MD2, MD4, and SHA-1.

\lead{\textsc{Decide}.} In every round, each revision's target tasks are executed four fresh times with both the candidate and the anchor. A decision needs scored executions of the candidate on at least 85\% of the training tasks. A small content step executes the candidate only on the tasks it can affect, together with a guard sample of twelve other tasks, and on every training task when the guard shows a change. A revision that makes the index more than half again as long must be clearly positive rather than a gain.

\lead{\textsc{Select}.} For the selection criterion (\S\ref{sec:final}), the agent executes each validation task twice with every kept library. When the best of them is not the latest, both get ten executions on each validation task before the bootstrap comparison. The run on CWEval, which has no validation split, returns its latest kept library.

\lead{Cost.} Evolution is paid once per agent and model, and its library also serves other agents and models. Per main run, the coding agent made 35,600 to 46,400 executions with 0.31 to 0.39M input tokens each. Its executions issued 496,000 model calls in Codex's run, 322,000 in Claude Code's, and 473,000 in Qwen Code's. \ProjectName{}'s own model calls, which analyze failures and plan and draft revisions, made 30,800, 27,900, and 33,200 calls in the same runs, 5.8, 8.0, and 6.6\% of all model calls; by the length of their transcripts, their input is about 0.6B tokens per run, 4 to 5\% of the coding agent's. At deployment, the library adds no model calls, only reading. On SeCodePLT's test tasks, an execution reads 0.11 (Codex), 0.15 (Claude Code), and 0.16M (Qwen Code) input tokens without guidance and 0.25 to 0.30M with the final library.

\lead{The floor on the paired spread.}\label{app:stats}
In Eq.~\ref{eq:paired}, the spread $\hat\sigma_d$ is floored at $0.42/\sqrt{n}$, where $0.42 = \sqrt{2 \cdot 0.1 \cdot
0.9}$ is the standard deviation of the difference of two independent outcomes at a 10\% rate of functional and secure
code, and $n$ is the number of executions per task the floor counts, since a task's outcome is then its mean over $n$
executions. Equivalently, $\mathrm{se} = \max\big(\hat\sigma_d/\sqrt{|T|},\ \sqrt{0.18/(n|T|)}\big)$. With $n = 1$, as in
every function-level run, this is the floor of one outcome per task. The repository-level run uses $n = 4$ for rounds on
the target tasks, which get four executions per task (\S\ref{sec:repo}). Without it, a few target tasks on which both
libraries happen to agree would give a standard error near zero and a gain that looks decisive.

\lead{The standard error of the stratified estimate.}
The two parts of Eq.~\ref{eq:strat} come from different executions, so their errors add by their weights:
$\mathrm{se}_{\mathrm{all}}^2 = w^2\,\mathrm{se}_{T_\delta}^2 + (1-w)^2\,\mathrm{se}_{\Ttr\setminus T_\delta}^2$, with
$\mathrm{se}_{T_\delta}$ the pooled standard error of Eq.~\ref{eq:pool}.

\lead{The retirement criterion.}
A revision in the ledger retires when its pooled gain, on $J$ or on $S_{|P}$, is negative with 95\% confidence, $\bar\Delta_{\delta} + 1.645\,\mathrm{se}_{\delta} < 0$, or after three rounds in which it showed a clear gain on neither, $\bar\Delta_{\delta} - 1.645\,\mathrm{se}_{\delta} \le 0$. A revision of an earlier entry counts its parent's rounds, so an idea that fails in every form retires as one.

\lead{Records.} Every step writes a plain-language journal entry and a page per phase, from which the account in Appendix~\ref{app:step} is taken.

\end{document}